\documentclass[prd,twocolumn]{revtex4}
\usepackage{bm}
\usepackage{epsfig}
\usepackage{amsmath}
\newcommand{\be}{\begin{equation}}
\newcommand{\ee}{\end{equation}}
\newcommand{\ba}{\begin{eqnarray}}
\newcommand{\ea}{\end{eqnarray}}
\newcommand{\bml}{\begin{mathletters}}
\newcommand{\eml}{\end{mathletters}}
\newcommand{\bes}{\begin{subequations}}
\newcommand{\ees}{\end{subequations}}

\newcommand{\bi}{\begin{itemize}}
\newcommand{\ei}{\end{itemize}}

\begin{document}
\title{A Model of Dark Energy and Dark Matter}
\author{P.Q. Hung}
\email[]{pqh@virginia.edu}
\affiliation{Dept. of Physics, University of Virginia, \\
382 McCormick Road, P. O. Box 400714, Charlottesville, Virginia 22904-4714, 
USA}
\date{\today}
\begin{abstract}
A dynamical model for the dark energy is presented in which the
``quintessence'' field is the axion, $a_Z$, of a spontaneously broken
global $U(1)_{A}^{(Z)}$ symmetry whose potential is induced by the
instantons of a new gauge group $SU(2)_Z$. The $SU(2)_Z$ coupling
becomes large at a scale $\Lambda_Z \sim 10^{-3}\,eV$ starting
from an initial value $M$ at high energy which is of the order of the
Standard Model (SM) couplings at the same scale $M$. A perspective on
a possible unification of $SU(2)_Z$ with the SM will be
briefly discussed. We present a scenario in which $a_Z$
is trapped
in a false vacuum characterized by an energy density $\sim
(10^{-3}\,eV)^4$. The lifetime of this false vacuum is estimated
to be extremely large. Other estimates relevant to
the ``coincidence issue'' include the ages of the
universe when the $a_Z$ potential became effective, when
the acceleration ``began'' and when
the energy density of the false vacuum became comparable
to that of (baryonic and non-baryonic) matter.
Other cosmological consequences include
a possible candidate for the weakly interacting (WIMP)
Cold Dark Matter as well as a scenario for leptogenesis. 
A brief discussion on possible
laboratory detections of some of the particles contained in the model
will also be presented. 

\end{abstract}
\maketitle

\section{Introduction}

The nature of the dark energy 
(responsible for an accelerating
universe \cite{acceleration}) is one of the 
deepest problem in contemporary cosmology.
Supernovae observations at redshifts $1.25 \leq z \leq 1.7$ when combined
with cosmic microwave background (CMB) and cluster data 
gave an equation of state $w=p/\rho = -1.02+0.13-0.19$ 
\cite{riess} and are
consistent with a generic $\Lambda\,CDM$ model where  $w=-1$
independently of $z$. 
Most recently, distance measurements of 71 high redshift Type Ia
supernovae by the Supernova Legacy Survey (SNLS) up to $z=1$ combined with
measurements of baryon acoustic oscillations by the Sloan Digital
Sky Survey also fits a flat $\Lambda\,CDM$ with constant $w = -1.023 \pm
0.090 \pm 0.054$ \cite{snls}.
Future proposed measurements to test whether or not $w$ is time-varying
will be of crucial importance.
Various forms of Quintessence had been proposed to describe the 
present accelerating universe \cite{sahni}. 
A generic feature of these models is
the presence of a time-varying $w$. However, it is a known fact 
that the dark
energy is subdominant at higher values of redshift which makes
it much harder to detect the $z$-dependence of $w$ \cite{doran}. Until
this is resolved, it is practically impossible to distinguish
the class of quintessence models with time-varying $w$ from one
in which $w$ is practically constant and is equal to $-1$.
However, one should keep in mind that several quintessence
models typically predict $w>-1$ now with many of them
having $w \agt -0.8$. In fact, one can try to reconstruct
the quintessence potential as had been done recently by \cite{sahlen}
whose analysis of recent data appeared to favor a cosmological
constant.

Is there a quintessence scenario in which $w=-1$ for a large range
of $z$ and which mimics the $\Lambda\,CDM$ model? Can such
a scenario make predictions that go beyond the accelerating universe
issue and that can be tested experimentally?
These are the types of questions we wish to address in this paper.

There exists a well-known phenomenon that can be readily applied
to the search for models that mimic $\Lambda\,CDM$: The
idea of the false vacuum. It has been used in the construction
of models of early inflation (although a ``standard model'' is
yet to be found) \cite{kolb}. In its simplest version, the potential of
a scalar field (whose nature depends on a given model) develops
two local minima: a ``false'' and a ``true'' one, as the
temperature drops below a certain critical value. In this class of 
models, the universe is trapped in the false vacuum and the
total energy density of the universe is soon dominated by the
energy of this false vacuum, leading to an exponential
expansion. For the early inflation case, models have been constructed
to deal with the so-called graceful exit problem, i.e. how to go
from the false vacuum to the true vacuum without creating gross 
inhomogeneities, resulting in a class of so-called new inflationary
scenarios (see \cite{kolb} for an extensive list of references).

Is the fact that present measurements appear to be consistent with
a flat $\Lambda\,CDM$ model with a constant $w=-1$ an
indication that we have been and are still living in a false vacuum
with an energy density $\rho_{vac} \sim (10^{-3}\,eV)^4$?
If that is the case, when did we get trapped in that false vacuum
and when are we getting out of it? And where does this false vacuum
come from?

In this paper, we would like to explore the above 
possibility and present a model for the false vacuum scenario.
First, we will postulate the existence of an
unbroken gauge group $SU(2)_Z$ \cite{su2} 
and show that, starting with a gauge
coupling comparable in value to the Standard Model (SM) couplings
at some high energy scale ($\sim 10^{16}\,GeV$), it becomes strongly
interacting at a scale $\sim 10^{-3}\,eV$. This new gauge group $SU(2)_Z$
\cite{zophos} can be seen to come from the breaking $E_6$ \cite{bj} into
$SU(2)_Z \otimes SU(6)$, where $SU(6)$ can, as one possible
scenario, first break down
to $SU(3)_c \otimes SU(3)_L \otimes U(1)$ and then to
$SU(3)_c \otimes SU(2)_L \otimes U(1)_Y$,
the details of which will be dealt with in a separate
paper \cite{hung2}. Next, we will list the particle content of our model
and present an argument showing how the $SU(2)_Z$ instanton-induced
axion potential can provide a model
for the aforementioned false vacuum with the
desired energy density \cite{jain}. We will compute
the transition rate to the true vacuum and show that it is
plausible that the universe was trapped in this false vacuum 
and will be accelerating for a very, very long time. We then
show that the particle spectrum of the model contains
fermions which have the necessary characteristics of
being candidates for a WIMP Cold Dark Matter. Finally,
we will briefly discuss the possibility of SM leptogenesis
in our model where the SM lepton number violation
comes from the asymmetry in the decay of a ``messenger''
scalar field which carries both $SU(2)_Z$ and $SU(2)_L$
quantum numbers. A more detailed version of this leptogenesis
scenario will appear in a separate paper \cite{hung3}. We will end with
a brief discussion of the possibility of detection
for the messenger field and the $SU(2)_Z$ fermions (CDM candidates).

\section{$SU(2)_Z$  as a new strong intersection at
extremely low energy}

In this section, we would like to discuss the possibility
of a new asymtotically-free gauge group, $SU(2)_Z$, which can grow
strong at an extremely low energy scale such as $\sim
10^{-3}\,eV$, starting with a coupling of the same
order as the Standard Model couplings at high energies
and, in particular, at some ``GUT'' scale $\sim
10^{16}\,GeV$ \cite{su2}. We first show
how, using the particle content of the model, the
$SU(2)_Z$ gauge coupling evolves from an initial value which
is close to those of the SM couplings at a typical
GUT scale $M$ to $\alpha_Z = g_{Z}^2/4\,\pi \sim 1$ at a
scale $\Lambda_Z \sim 10^{-3}\,eV$. (In a GUT scenario like
the $E_6$ example mentioned above, $\Lambda_Z$ could be
seen as being generated from the GUT scale $M$.)
Turning things around, one can ask the
following question: If one would like to have 
$\alpha_Z = g_{Z}^2/4\,\pi \sim 1$ at a
scale $\sim 10^{-3}\,eV$, what should the initial value of 
$\alpha_Z$ be at high energies in order for this condition
to be fulfilled? As we shall see below, it turns out that
this initial value is correlated with 
the $SU(2)_Z$ particle content and on the masses of
the SM-singlet $SU(2)_Z$ fermions in an interesting way:
$\alpha_Z(M)$ {\em decreases} as the masses of the 
$SU(2)_Z$ fermions {\em increase}. If we wish $\alpha_Z(M)$
to be close in value to the SM couplings, we find the
masses of these fermions to be in the $GeV$ region,
an interesting range for the dark matter as we shall see below.
We will then discuss a possible
origin of $SU(2)_Z$ from a grand unified point of view with more
details to be presented elsewhere \cite{hung2}.

\subsection{The $SU(2)_Z$ model and its particle content}
\label{content}

The gauge group that we are concerned with is

\be
\label{group}
G_{SM} \otimes SU(2)_Z\,,
\ee
where
\be
G_{SM} = SU(3) \otimes SU(3)_L \otimes U(1)_Y\,.
\ee
The $SU(2)_Z$ particle content is as follows.
\bi

\item Two fermions: $\psi^{(Z)}_{(L,R),i} = (1,1,0,3)$ 
under $ SU(3) \otimes SU(2)_L \otimes U(1)_Y \otimes SU(2)_Z$,
where $i=1,2$. The reasons for having two such fermions
will be made clear below when we discuss the evolution
of the $SU(2)_Z$ gauge coupling.

\item Messenger scalar fields: 
$\varphi^{(Z)} =(\varphi^{(Z),0},\varphi^{(Z),-})=
(1,2,Y_{\varphi}=-1,\,2)$ or two
${\tilde{\bm{\varphi}}}_{i}^{(Z)} =({\tilde{\bm{\varphi}}}^{(Z),0},
{\tilde{\bm{\varphi}}}
^{(Z),-}_{i})=
(1,2,Y_{\tilde{\varphi}}=-1,\,3)$ 
under $SU(3) \otimes SU(2)_L
\otimes U(1)_Y \otimes SU(2)_Z$, where $i=1,2$. Again the reason
for having two ${\tilde{\bm{\varphi}}}$
(one of which will be assumed to be much heavier than the other)
will be made clear below.
Briefly spaeking, it has to do with the SM leptogenesis mechanism
proposed at the end of the manuscript. For this reason, the scenario
with two ${\tilde{\bm{\varphi}}}$ is more attractive than that
in which one has only a $SU(2)_Z$ doublet $\varphi^{(Z)}$. We discuss
both cases in the section on the gauge coupling evolution for
completeness and for the purpose of comparison.

\item Complex singlet scalar field: $\phi_{Z} =(1,1,0,1)$.

\ei

Finally, the SM particles are assumed to be singlets under $SU(2)_Z$,
namely
\bi

\item $q_L = (3,2,1/3,1)$; $u_R=(3,1,2/3,1)$; $d_R=(3,1,-1/3,1)$.

\item $l_L = (1,2,-1,1)$; $e_R = (1,1,-2,1)$.

\ei
The above notations are meant to be generic for each SM family.
We do not list the right-handed neutrinos, which we believe to
exist, since they are singlets under $G_{SM} \otimes SU(2)_Z$
and are not relevant for the present analysis. 
 
A few words are in order concerning the above choices. The fermions
are chosen to be triplets of $SU(2)_Z$ in order to ``slow down'
the evolution of the $SU(2)_Z$ coupling. The messenger scalar
fields are chosen for two purposes: 1) to contribute to the $\beta$ 
function of $SU(2)_Z$ and 2) to connect the aforementioned fermions
to their SM counterparts. The singlet complex scalar field is
introduced in the manner of Peccei-Quinn \cite{PQ}. 
The instanton-induced ``axion-potential'' is used to model 
the dark energy, as we shall see below.

\subsection{The Lagrangian of the model}
\label{lag}

The Lagrangian for $G_{SM} \otimes SU(2)_Z$ is
\ba
\label{lagrangian}
{\cal L}&=& {\cal L}_{SM} + {\cal L}^{Z}_{kin} + {\cal L}_{yuk}
+ {\cal L}_{CP} -V(|\tilde{\bm{\varphi}}^{(Z)}|^2
\; {\rm or}\; |\varphi^{(Z)}|^2)
\nonumber \\
& &-V(|\phi_{Z}|^2)\,,
\ea
where ${\cal L}_{SM}$ is the well-known SM Lagangian, which does
not need to be explicitely written down here, and where
\ba
\label{Zkin}
{\cal L}^{Z}_{kin}& =& -\frac{1}{4} {\bf G}_{\mu\nu}^{(Z)}.
{\bf G}^{(Z),\mu\nu} + (\sum_{i} \frac{1}{2} (D_{\mu}\,
\tilde{\bm{\varphi}}_{i}^{(Z)})^{\dag}.(D^{\mu}\,
\tilde{\bm{\varphi}}_{i}^{(Z)})\,   \nonumber \\
& &{\rm or}\; \frac{1}{2} (D_{\mu}\,
\varphi^{(Z)})^{\dag}(D^{\mu}\,
\varphi^{(Z)})) \nonumber \\
& & + \sum_{i} i \bar{\psi}^{(Z)}_{(L,R),i}\not\!D
\psi^{(Z)}_{(L,R),i}\,,
\ea
\ba
\label{yuk}
{\cal L}_{yuk}&= & \sum_{i}\,\sum_{m}( g_{\tilde{\varphi}_{1}\,m}^{i}\,
\bar{l}_{L}^{m}\,
\tilde{\bm{\varphi}}_{1}^{(Z)}\,\psi^{(Z)}_{i,R}+ 
g_{\tilde{\varphi}_{2}\,m}^{i}\,\bar{l}_{L}^{m}\,
\tilde{\bm{\varphi}}_{2}^{(Z)}\,\psi^{(Z)}_{i,R}) \nonumber \\
&& +\sum_{i} K_{i} \, \bar{\psi}^{(Z)}_{L,i}\,\psi^{(Z)}_{R,i}\,\phi_{Z}
+ h.c. \,,
\ea
\be
\label{cp}
{\cal L}_{CP} = \frac{\theta_Z}{32\,\pi^2} {\bf G}_{\mu\nu}^{(Z)}.
\tilde{{\bf G}}^{(Z),\mu\nu}\,.
\ee
The covariant derivative acting on $\varphi$ is given by
\bes
\label{covariant1}
\be
D_{\mu}\varphi^{(Z)} = (\partial_{\mu} -i g \frac{\bm{\tau}}{2}.
{\bf W}_{\mu} +i \frac{g^{'}}{2} B_{\mu} -i g_Z \frac{\bm{\tau}}{2}.
{\bf A}^{(Z)}_{\mu})\varphi^{(Z)}\,,
\ee
\be
D_{\mu}\tilde{\bm{\varphi}}_{i}^{(Z)}= ((\partial_{\mu} -
i g \frac{\bm{\tau}}{2}.
{\bf W}_{\mu} +i \frac{g^{'}}{2} B_{\mu} -i g_Z {\bf T}.
{\bf A}^{(Z)}_{\mu})\tilde{\bm{\varphi}}_{i}^{(Z)}\,,
\ee
\ees
and that acting on $\psi^{(Z)}_{(L,R),i}$ is given by
\be
\label{covariant2}
D_{\mu}\psi^{(Z)}_{(L,R),i} = (\partial_{\mu} -i g_Z {\bf T}.
{\bf A}^{(Z)}_{\mu}) \psi^{(Z)}_{(L,R),i}\,,
\ee
where $(T^{i})_{jk} =i\, \epsilon^{ijk}$. 
In Eqs. (\ref{Zkin},\ref{cp}, \ref{covariant1}, \ref{covariant2}), 
we use boldfaces to express explicitely the triplet nature
of the $SU(2)_Z$ gauge fields and $\tilde{\varphi}$. Also,
in Eq. (\ref{Zkin},\ref{yuk}), the sum over $m$ means that we are
summing over the number of SM families while the sum
over $i$ means that we are summing over the two $SU(2)_Z$
fermions and the two triplet scalars. The coefficients
$g_{\tilde{\varphi}_{1}\,m}$, $g_{\tilde{\varphi}_{2}\,m}$,
and $K_i$ are, in general, {\em complex}.

\subsection{Global symmetries}
\label{global}

The Lagrangian written above exhibits a $U(1)_{A}^{(Z)}$ global symmetry.
In fact, Eqs.(\ref{yuk},\ref{cp}) are invariant
under the following $U(1)_{A}^{(Z)}$ phase transformation:
\bes
\be
\label{phase0}
\psi^{(Z)}_{i} \rightarrow e^{i\alpha \gamma_{5}}\,\psi^{(Z)}_{i}\,,
\ee
\be
\label{phase1}
\psi^{(Z)}_{L,i} \rightarrow e^{-i\alpha}\,\psi^{(Z)}_{L,i} \,,
\ee
\be
\label{phase2}
\psi^{(Z)}_{R,i} \rightarrow e^{i\alpha}\,\psi^{(Z)}_{R,i} \,,
\ee
\be
\label{phase3}
\phi_{Z} \rightarrow e^{-2i\alpha}\,\phi_{Z} \,,
\ee
\be
\label{phase4}
\theta_Z \rightarrow \theta_Z - 4\,\alpha \,,
\ee
\be
\label{phase5}
l_{L}^{m} \rightarrow e^{i\alpha}\,l_{L}^{m}\,,
\ee
\be
\label{phase6}
\tilde{\bm{\varphi}}_{i}^{(Z)} \rightarrow \tilde{\bm{\varphi}}_{i}^{(Z)}\,.
\ee
\ees

Since ${\cal L}_{SM}$ contains Yukawa couplings between the SM leptons
to the SM Higgs fiels $\phi_{SM}$ of the form $\bar{l}_{L}^{m}\,
\phi_{SM}\,l_{R}^{n}$ (and also $\bar{l}_{L}^{m}\,
\tilde{\phi}_{SM}\,\nu_{R}^{n}$ for the neutral leptons), 
where $l_{R}^{n}$ ($\nu_{R}^{n}$) denotes the charged (neutral)
right-handed leptons, it will be invariant under the above
$U(1)_{A}^{(Z)}$ global symmetry provided
\be
\label{phaser}
l_{R}^{m}\; (\nu_{R}^{m}) \rightarrow e^{i\alpha}\,l_{R}^{m}\;
(\nu_{R}^{m})\,,
\ee
when we use the transformation (\ref{phase5}). All other SM particles
are unchanged under $U(1)_{A}^{(Z)}$.

The above $U(1)_{A}^{(Z)}$ symmetry plays an important role in the
emergence of an $SU(2)_Z$ instanton-induced axion potential
which could drive the present accelerating universe,
as we shall see below.

\subsection{Spontaneous breadown of $U(1)_{A}^{(Z)}$ and
masses of $\psi^{(Z)}_{1,2}$}
\label{psizmass}

In this section, we will discuss the masses of particles,
$\psi^{(Z)}_{1,2}$ and $\tilde{\bm{\varphi}}^{(Z),\dag}_{1,2}$
or $\varphi^{(Z)}$, which carry $SU(2)_Z$ quantum numbers since
we would like to examine the evolution of the $SU(2)_Z$
gauge coupling. This in turn will put interesting constraints
on these masses. Those of $\psi^{(Z)}_{1,2}$ come from the
spontaneous breaking of $U(1)_{A}^{(Z)}$ described above, while the
scalar masses are arbitray gauge-invariant parameters.

The spontaneous breakdown of $U(1)_{A}^{(Z)}$ gives masses to
$\psi^{(Z)}_{i}$ through Eq. (\ref{yuk}). With the potential
$V(\phi_{Z}^{\dag}\, \phi_{Z})$ of the form
\be
\label{potential1}
V(\phi_{Z}^{\dag}\, \phi_{Z})= (\lambda/4)\,(\phi_{Z}^{\dag}\, \phi_{Z}
-v_{Z}^2)^{2}\,, 
\ee
the vacuum-expectation-value (VEV) of
$\phi_{Z}$ is given by
\be
\label{VEV}
\langle \phi_{Z} \rangle = v_Z \,.
\ee
where $v_Z$ is real. 
In fact, one can write
\be
\label{phiz}
\phi_{Z} = v_{Z}\,\exp(ia_Z/v_{Z}) + \sigma_Z \,,
\ee
where $\langle \sigma_Z \rangle =0$ and
$\langle a_Z \rangle =0$. The field $a_Z$,
the $SU(2)_Z$ axion, would be a massless Nambu-Goldstone boson 
if it were not
for the fact that the $U(1)_{A}^{(Z)}$ symmetry is explicitely
broken by the $SU(2)_Z$ gauge anomaly which we will discuss
in Section ().

There is a remaining unbroken $Z(2)$ (for two flavors)
symmetry of $U(1)_{A}^{(Z)}$. This implies that there are
two degenerate vacua. In a similar fashion to \cite{sikivie},
we will add a soft breaking term to $U(1)_{A}^{(Z)}$ to
lift this degeneracy. This has an important implication
to the dark energy scenario discussed below.

Before discussing the masses of $\psi^{(Z)}_{i}$, an important
remark should be pointed out. Since we would like $SU(2)_Z$
to be {\em unbroken}, one can choose 
$V(\tilde{\bm{\varphi}}^{(Z),\dag}.
\tilde{\bm{\varphi}}^{(Z)} \; {\rm or}\; \varphi^{(Z),
\dag}\,\varphi^{(Z)})$ such that $\tilde{\bm{\varphi}}^{(Z)}$
or $\varphi^{(Z)}$ has vanishing vacuum expectation value.
Therefore, with the triplet ($\tilde{\bm{\varphi}}^{(Z)}$)
scenario, Eq. (\ref{yuk}) {\em does not} give a mass mixing
between $\psi^{(Z)}_{i}$ and the SM leptons. The masses
of $\psi^{(Z)}_{i}$ comes from their couplings to $\phi_{Z}$.

From Eqs. (\ref{yuk}), one obtains
\bes
\label{masses}
\be
m_{\psi^{(Z)}_{1}} = |K_1| v_Z \,,
\ee
\be
m_{\psi^{(Z)}_{2}} = |K_2| v_Z \,.
\ee
\ees

In Section (\ref{evolution}) where we discuss the evolution of the $SU(2)_Z$
gauge coupling, it will be seen how one can obtain constraints
on $m_{\psi^{(Z)}_{1,2}}$ and hence on $|K_{1,2}| v_Z$. As we shall
see below in both the sections on the evolution of the $SU(2)_Z$
coupling as well as the section on dark matter, one expects
at least $m_{\psi^{(Z)}_{2}}$ to be around $100\,GeV$ or so which
implies that $v_Z$ could range in the several hundreds of GeVs.

\subsection{Masses of the messenger scalar fields}
\label{messenger}

The other particles which enter the evolution 
of the $SU(2)_Z$ gauge coupling
at one-loop are $\tilde{\bm{\varphi}}_{1,2}^{(Z)}$
or $\varphi^{(Z)}$. It is well-known
that the masses in the scalar sector represent a notoriously
difficult problem to tackle, in particular the so-called
gauge hierarchy problem which is present when there exists several
widely different mass scales in the model, 
e.g. $\Lambda_{EW}$ and $\Lambda_{GUT}$. There exists a continuing large
body of works on the subject with the essential points being as follows.
First, there is a fine-tuning problem already at the tree level that sets
the small and large scales apart. Second, the tree-level fine-tuning
can get spoiled by radiative corrections. Supersymmetry provides
an elegant candidate for making this second problem ``technically
natural''. Other alternative attempts have been made to keep the
``small scale'' radiatively stable. Our model falls into the same
category as a typical GUT scenario which is usually characterized
by two sets of widely different scales such as $\Lambda_{EW}$ 
and $\Lambda_{GUT}$. It is beyond the scope of this paper to
get into the (more general) gauge hierarchy problem and we
will restrict to a discussion of how masses are obtained at
the tree-level. We will assume, as with a generic GUT scenario,
that the ``small scale'' is radiatively stable by either
supersymmetry or some other mechanisms.

As we have
mentioned above, the scalar fields which carry both SM and $SU(2)_Z$
quantum numbers, are assumed to have zero vacuum expectation values
in order for $SU(2)_Z$ to be unbroken. The potential
$V(\tilde{\bm{\varphi}}^{(Z),\dag}.
\tilde{\bm{\varphi}}^{(Z)} \; {\rm or}\; \varphi^{(Z),\dag}\,\varphi^{(Z)})$
will the contain a gauge-invariant mass term of the form:
\be
\label{phitilde}
\sum_{i=1}^{2} \frac{1}{2}\,m_{\tilde{\bm{\varphi}}^{(Z)},i}^{0,2}\,
\tilde{\bm{\varphi}}^{(Z),\dag}_{i}.
\tilde{\bm{\varphi}}^{(Z)}_{i}  \,,
\ee
or
\be
\label{phi}
\frac{1}{2}\,m_{\varphi^{(Z)}}^{0,2}\,
\varphi^{(Z),\dag}\,\varphi^{(Z)} \,.
\ee

In addition to the above ``bare'' masses, the messenger fields can acquire
masses by possible couplings to scalars that do have non-vanishing VEVs
such as $\phi_{Z}$ and $\phi_{SM}$, and possible other scalar fields
$\phi_j$ which can come from the GUT sector as we will see below.
We can have
\ba
\label{cross1}
{\cal L}_{\tilde{\bm{\varphi}}} &=& \sum_{i}\,\tilde{\bm{\varphi}}^{(Z),\dag}_{i}.
\tilde{\bm{\varphi}}^{(Z)}_{i}(\tilde{\lambda}_{iZ}\,
\phi_{Z}^{\dag}\, \phi_{Z} + \tilde{\lambda}_{iSM}\,
\phi_{SM}^{\dag}\, \phi_{SM} + \nonumber \\ 
&&\sum_{j}\,\tilde{\lambda}_{i,j}
\phi_{j}^{\dag}\, \phi_{j})\,,
\ea
or
\be
\label{cross2}
{\cal L}_{\varphi} = \varphi^{(Z),\dag}\,\varphi^{(Z)}
(\lambda_{Z}\,\phi_{Z}^{\dag}\, \phi_{Z} + \lambda_{SM}\,
\phi_{SM}^{\dag}\, \phi_{SM} + \sum_{j}\,\lambda_{j}
\phi_{j}^{\dag}\, \phi_{j})\,,
\ee
where
\be
\label{vev2}
\langle \phi_{Z} \rangle = v_Z\,;\,
\langle \phi_{SM} \rangle = v_{SM}\,;\,
\langle \phi_{j} \rangle = v_j\,.
\ee
and where we will assume
\be
\label{vev3}
v_j \gg v_Z \sim O(v_{SM})\,.
\ee

The effective mass squared can be now written as
\ba
\label{mass1}
m_{\tilde{\bm{\varphi}}^{(Z)},(1,2)}^{eff,2}&=&
m_{\tilde{\bm{\varphi}}^{(Z)},(1,2)}^{0,2} + 2 \tilde{\lambda}_{(1,2)Z}v_{Z}^2
+2 \tilde{\lambda}_{(1,2)SM}v_{SM}^2 \nonumber \\ 
&&+ \sum_{j} 2\,\tilde{\lambda}_{(1,2),j}
v_{j}^2 \,,
\ea
or
\ba
\label{mass2}
m_{\varphi^{(Z)}}^{eff,2} &=& 
m_{\varphi^{(Z)}}^{0,2} + 2 \lambda_{(Z}v_{Z}^2
+2 \lambda_{SM}v_{SM}^2 \nonumber \\
&&+ \sum_{j} 2\,\lambda_{j}
v_{j}^2 \,.
\ea

As we have mentioned in Section (\ref{content}), the scenario
with two ${\tilde{\bm{\varphi}}}$ is more attractive than that
in which one has only a $SU(2)_Z$ doublet $\varphi^{(Z)}$ because
of the leptogenesis scenario proposed at the end of the manuscript.
However, for completeness, we will discuss both cases in this section
in order to compare them in the section on the evolution of
the $SU(2)_Z$ gauge coupling. As we shall see in that section, one
of the two ${\tilde{\bm{\varphi}}}$s will be required to be much
heavier ( mass of O(``GUT'') scale) than the other, 
of mass of $O(\Lambda_{EW})$, in order for
the initial high energy value of the $SU(2)_Z$ coupling to be
of the order of the SM couplings.

\bi

\item Let us first discuss the triplet $\tilde{\bm{\varphi}}^{(Z)}$ case.
From Section () on the RG evolution of the $SU(2)_Z$ coupling, we will see
that one needs $m_{\tilde{\bm{\varphi}}^{(Z)},1}^{eff,2} \sim
O(\Lambda_{EW}^2)$ and $m_{\tilde{\bm{\varphi}}^{(Z)},2}^{eff,2} \sim
O(\Lambda_{GUT}^2)$, i.e. $m_{\tilde{\bm{\varphi}}^{(Z)},1}^{eff}
\ll m_{\tilde{\bm{\varphi}}^{(Z)},2}^{eff}$.
Since $2 \tilde{\lambda}_{(1,2)Z}v_{Z}^2
+2 \tilde{\lambda}_{(1,2)SM}v_{SM}^2  \sim O(\Lambda_{EW})$, 
where $\Lambda_{EW}$ is the electroweak scale, one would then require
\be
\label{constraint1}
m_{\tilde{\bm{\varphi}}^{(Z)},1}^{0,2}+ \sum_{j} 2\,\tilde{\lambda}_{1,j}
\,v_{j}^2 \leq O(\Lambda_{EW}^2) \,,
\ee
if we wish to have $m_{\tilde{\bm{\varphi}}^{(Z)},1}^{eff,2} \sim
O(\Lambda_{EW}^2)$. The constraint (\ref{constraint1}) coupled
with Eq. (\ref{mass1}) would guarantee that 
$m_{\tilde{\bm{\varphi}}^{(Z)},2}^{eff,2} \sim O(v_j^2)$
provided $\tilde{\lambda}_{2,j}\,>\,\tilde{\lambda}_{1,j}$. Some
cautionary words concerning the above constraint will be mentioned
at the end of this section.

From (\ref{constraint1}), one could entertain several possibilities.
The most obvious is one in which $\tilde{\lambda}_{1,j}=0;
\tilde{\lambda}_{2,j}> 0$ and
$m_{\tilde{\bm{\varphi}}^{(Z)},1}^{0,2} \sim
m_{\tilde{\bm{\varphi}}^{(Z)},2}^{0,2} \sim
O(\Lambda_{EW}^2)$. This will guarantee, at tree-level, that
$m_{\tilde{\bm{\varphi}}^{(Z)},1}^{eff,2} \sim
O(\Lambda_{EW}^2)$ and $m_{\tilde{\bm{\varphi}}^{(Z)},2}^{eff,2} 
\sim O(v_j^2) \gg m_{\tilde{\bm{\varphi}}^{(Z)},1}$.

Another possibility is one in which one assumes a global
$SU(2)$ symmetry among $\tilde{\bm{\varphi}}_{1}^{(Z)}$ and
$\tilde{\bm{\varphi}}_{2}^{(Z)}$, which, for simplicity, will
be denoted by $SU(2)_{\tilde{\bm{\varphi}}}$. The doublet
of $SU(2)_{\tilde{\bm{\varphi}}}$ is now
\be
\label{globsu2}
\tilde{\bm{\varphi}} = 
\left(
\begin{array}{c}
\tilde{\bm{\varphi}}_{2}^{(Z)} \\
\tilde{\bm{\varphi}}_{1}^{(Z)}
\end{array} \right)\,.
\ee
A $SU(2)_{\tilde{\bm{\varphi}}}$-invariant term, including
``bare'' masses, can be written as
\ba
\label{bare}
&(\frac{1}{2}\,m_{\tilde{\bm{\varphi}}^{(Z)}}^{0,2}+
\tilde{\lambda}_{iZ}\,
\phi_{Z}^{\dag}\, \phi_{Z} + \tilde{\lambda}_{iSM}\,
\phi_{SM}^{\dag}\, \phi_{SM})\, 
\tilde{\bm{\varphi}}^{\dag}\,\tilde{\bm{\varphi}} =& \nonumber \\
& (\frac{1}{2}\,m_{\tilde{\bm{\varphi}}^{(Z)}}^{0,2}+
\tilde{\lambda}_{iZ}\,
\phi_{Z}^{\dag}\, \phi_{Z} + \tilde{\lambda}_{iSM}\,
\phi_{SM}^{\dag}\, \phi_{SM}) &  \nonumber \\
& \times (\sum_{i=1}^{2} \tilde{\bm{\varphi}}^{(Z),\dag}_{i}.
\tilde{\bm{\varphi}}^{(Z)}_{i}) & \,.
\ea
One can have an {\em explicit} $SU(2)_{\tilde{\bm{\varphi}}}$-breaking
term in the coupling of $\tilde{\bm{\varphi}}$ to $\phi_j$ as
follows
\ba
\label{breaking}
&\tilde{\bm{\varphi}}^{\dag}\,\tau_3 \,\tilde{\bm{\varphi}}
(\sum_{j}\,\tilde{\lambda}_{j}
\phi_{j}^{\dag}\, \phi_{j})=& \nonumber \\
& (\tilde{\bm{\varphi}}^{(Z),\dag}_{2}.
\tilde{\bm{\varphi}}^{(Z)}_{2}-\tilde{\bm{\varphi}}^{(Z),\dag}_{1}.
\tilde{\bm{\varphi}}^{(Z)}_{1}) \times
 (\sum_{j}\,\tilde{\lambda}_{j}
\phi_{j}^{\dag}\, \phi_{j})& \,.
\ea

With the VEVs given in Eq. (\ref{vev2}), one now obtains
\ba
\label{mass3}
m_{\tilde{\bm{\varphi}}^{(Z)},(1,2)}^{eff,2}&=&
m_{\tilde{\bm{\varphi}}^{(Z)}}^{0,2} + 2 \tilde{\lambda}_{Z}v_{Z}^2
+2 \tilde{\lambda}_{SM}v_{SM}^2 \nonumber \\ 
&&\mp \sum_{j} 2\,\tilde{\lambda}_{j}
v_{j}^2 \,,
\ea
From Eq. (\ref{mass3}), one can see that the constraint (\ref{constraint1})
is now translated into
\be
\label{constraint2}
m_{\tilde{\bm{\varphi}}^{(Z)}}^{0,2}- \sum_{j} 2\,\tilde{\lambda}_{j}
\,v_{j}^2 \leq O(\Lambda_{EW}^2) \,,
\ee
in order for $m_{\tilde{\bm{\varphi}}^{(Z)},1}^{eff,2} \sim
O(\Lambda_{EW}^2)$. Furthermore, once the constraint (\ref{constraint2})
is satisfied, one automatically obtains 
$m_{\tilde{\bm{\varphi}}^{(Z)},2}^{eff,2} \sim O(v_j^2) \gg
m_{\tilde{\bm{\varphi}}^{(Z)},1}^{eff,2} \sim O(\Lambda_{EW}^2)$.
However, one needs a delicate cancellation in (\ref{constraint2}).
For this reason, it is not clear that this is more attractive
than the first possibility discussed above. The
purpose here is simply to mention various scenarios.

\item For the $SU(2)_Z$ doublet $\varphi^{(Z)}$, the discussion
of its mass is identical to the first possibility mentioned
above. In brief, if $\lambda_{j}=0$ and $m_{\varphi^{(Z)}}^{0,2}
\sim O(\Lambda_{EW}^2)$, one obtains at tree-level
$m_{\varphi^{(Z)}}^{eff,2} \sim O(\Lambda_{EW}^2)$.

\ei

As we have mentioned at the beginning of this section, it will be assumed
that there is a mechanism (supersymmetry, etc...) which will
make the smaller mass scale radiatively stable.
We will again see that the evolution of the $SU(2)_Z$
gauge coupling puts a non-trivial constraint on
$m_{\tilde{\bm{\varphi}}^{(Z)},i}$ or $m_{\varphi^{(Z)}}$.

\subsection{Evolution of the $SU(2)_Z$ gauge coupling}
\label{evolution}

In this section, we will study the evolution of the $SU(2)_Z$ gauge 
coupling with the particle content listed in Section (\ref{content}).
In particular, we will explore the conditions under which the
coupling, $\alpha_Z = g_Z^2/4\,\pi$, starting with an 
initial value close to that of the SM couplings at high energy 
(which would suggest some type of unification),
increases to $\alpha_Z \sim 1$ at $\Lambda_Z \sim 3 \times 10^{-3}\,eV$.
In this discussion, we will see how the initial value of the coupling
depends on the masses of the $SU(2)_Z$ particles if we require
that $\alpha_Z \sim 1$ at $\Lambda_Z \sim 3 \times 10^{-3}\,eV$.
For this analysis, we will use a two-loop $\beta_Z$ function to
study the evolution of $\alpha_Z$.

The $SU(2)_Z$ fermion masses are given by Eq. (\ref{masses}). 
Since both the Yukawa couplings and $v_Z$ are arbitrary, in
the following we will assume that the Yukawa couplings are small
enough so that we can neglect them in $\beta_Z$.

The evolution equation for $\alpha_Z$ at two loops can be written
as
\be
\label{RG}
\frac{\text{d}\alpha_Z}{\text{dt}} = -8\,\pi\,b_{Z}^{0}\,\alpha_{Z}^2
-32\,\pi^2\,b_{Z}^{1}\,\alpha_{Z}^3 \,,
\ee
where
\bes
\label{beta3}
\be
b_{Z}^{0} = (\frac{22}{3}-\frac{8}{3}\, n_{F}-\frac{4}{3}\, n_S)/
16\,\pi^2\,,
\ee
\be
b_{Z}^{1} = (\frac{4}{3})\,(34 - 32\,n_F - 28\,n_S)/(16\,\pi^2)^2 \,,
\ee
\ees
for the $SU(2)_Z$ triplet scalar case ${\tilde{\bm{\varphi}}}_{i}^{(Z)} 
= (3,2)$ (under $SU(2)_Z \otimes SU(2)_L$), and
\bes
\label{beta2}
\be
b_{Z}^{0} = (\frac{22}{3}-\frac{8}{3}\, n_{F}-\frac{1}{3}\, n_S)/
16\,\pi^2\,,
\ee
\be
b_{Z}^{1} = (\frac{4}{3})\,(34 - 32\,n_F - \frac{13}{4}\,n_S)/(16\,\pi^2)^2 \,,
\ee
\ees
for the $SU(2)_Z$ doublet case $\varphi^{(Z)} =(2,2)$. In (\ref{beta3}) and
(\ref{beta2}), we have already taken into account that both
${\tilde{\bm{\varphi}}}_{i}^{(Z)}$ and $\varphi^{(Z)}$ are doublets under
$SU(2)_L$.

We will divide the evolution of $\alpha_Z$ into four regions.

I) Between a ``GUT'' scale $M$ and the scalar mass 
$m_{\tilde{\bm{\varphi}}_{1}^{(Z)}}$ (or $m_{\varphi^{(Z)}}$):
$n_F =2$ and $n_S=1$.

In the scalar triplet case, we will assume that 
$m_{\tilde{\bm{\varphi}}_{1}^{(Z)}} \sim M$ and will not include
it the evolution equations. The rationale for this assumption
stems in part from the fact that we wish $SU(2)_Z$ to be asymptotically
free below $M$ and in part from the leptogenesis scenario
alluded to above.

II) Between $m_{\tilde{\bm{\varphi}}_{1}^{(Z)}}$ (or $m_{\varphi^{(Z)}}$)
and $m_{\psi^{(Z)}_{2}}$: $n_F=2$ and $n_S=0$.

III) Between $m_{\psi^{(Z)}_{2}}$ and $m_{\psi^{(Z)}_{1}}$:
$n_F=1$ and $n_S=0$.

IV) Between $m_{\psi^{(Z)}_{1}}$ and $\Lambda_Z$:
$n_F=0$ and $n_S=0$.

Starting with a value for $\alpha_Z(M)$, one can use Eq. (\ref{RG})
to evolve it through the four regions, with the condition that
$\alpha_Z(\Lambda_Z) =1$. With this condition, one can immediately
see how, for a given $\alpha_Z(M)$, the evolution depends on 
the various mass thresholds. One can also see, for a given set
of masses, what $\alpha_Z(M)$ should be in order for
$\alpha_Z(\Lambda_Z) =1$. Since an exhaustive analysis of these
dependences is outside the scope of this paper, we will show
a few typical examples for the purpose of illustration and for
the discussion of the dark energy and dark matter scenarios.

For definiteness, we will take $M = 2 \times 10^{16}\,GeV$ and
$\Lambda_Z = 3 \times 10^{-3}\,eV$. Since we wish to illustrate
the range of masses which is attractive from a phenomenological
viewpoint, we will also set $m_{\tilde{\bm{\varphi}}_{1}^{(Z)}}
= 300\,GeV$ (and similarly for $m_{\varphi^{(Z)}}$).
We solve Eq. (\ref{RG}) numerically.

First we set $m_{\psi^{(Z)}_{2}} = 100\; GeV$.
We then show in Table I and Figures (\ref{fig1},
\ref{fig2}, \ref{fig3}, \ref{fig4}, \ref{fig5})
the dependences of $\alpha_Z(M)$
on $m_{\psi^{(Z)}_{1}}$. We show graphs corresponding to
$m_{\psi^{(Z)}_{1}}= 50\;GeV\,,\, 10\;GeV\,,\, 1\;GeV\,,\, 1\;MeV\,,\,
1\;eV$ respectively.
We can clearly see that, as we lower
the value for $m_{\psi^{(Z)}_{1}}$, $\alpha_Z(M)$ also decreases. In fact,
$\alpha_Z(M)$ varies from $1/41.6$ to $1/30.17$ as $m_{\psi^{(Z)}_{1}}$
varies from $50\,GeV$ to $1\,eV$. From Figs. (\ref{fig1},
\ref{fig2}, \ref{fig3}, \ref{fig4}, \ref{fig5}), one notices that
$\alpha_Z(E)$ and $\alpha_Z^{-1}(E)$ are relatively flat until
$E$ reaches the mass of the lightest of the two fermions, namely
$m_{\psi^{(Z)}_{1}}$. They then steepen and reaches unity at 
$\Lambda_Z = 3 \times 10^{-3}\,eV$.

\begin{table}
\caption{\label{tab:table1} Correlations between $m_1$, $m_2$
and $\alpha_{Z}^{-1}(M)$ with $\Lambda_Z = 3 \times 10^{-3}\,eV$
and $M= 2 \times 10^{16}\,GeV$, for the $SU(2)_Z$
triplet and doublet messenger field respectively.
The constraint is $\alpha_Z(\Lambda_Z) =1$.} 
\begin{ruledtabular}
\begin{tabular}{ccccc} 
&$m_1$&$m_2$&$\alpha_{Z}^{-1}(M)$ \\ \hline
$m_{\tilde{\bm{\varphi}}_{1}^{(Z)}}
= 300\,GeV$&
$50\,GeV$&$100\,GeV$&$41.6$ \\
&$10\,GeV$&$100\,GeV$&$40.86$ \\
&$1\,GeV$&$100\,GeV$&$39.83$ \\
&$1\,MeV$&$100\,GeV$&$36.7$ \\
&$1\,eV$&$100\,GeV$&$30.17$ \\ 
&$100\,GeV$&$200\,GeV$&$42.2$ \\ 
&$50\,GeV$&$200\,GeV$&$41.9$ \\ \hline
$m_{\varphi^{(Z)}}=300\,GeV$&
$50\,GeV$&$100\,GeV$&$47$ \\
\end{tabular}
\end{ruledtabular}
\end{table}

For the purpose of seeing
how e.g. a 10 \% change in the initial $\alpha_Z(M)$ affects the
scale where $\alpha_Z$ reaches unity, we show in Fig. (\ref{fig6}) 
$\alpha_Z^{-1}(E)$
for the case $m_{\psi^{(Z)}_{1}}= 50\;GeV$ with $\alpha_Z(M)= 1/38$
instead of $1/41.6$ used in Fig. (\ref{fig1}) (a 10 \% change). We notice
that this corresponds to $\Lambda_{Z}^{new} \sim 19 \times
\Lambda_Z \sim 5.7 \times 10^{-2}\,eV$, a still {\em very small} scale.
As we have mentioned above, for a given value of $\alpha_Z(M)$, one can
always choose $m_{\psi^{(Z)}_{1}}$ so that $\alpha_Z =1$ at
$\Lambda_{Z}$. (Also, without showing a plot, we find that, for the
same set of mass parameters, the choice of $\alpha_Z(M) =1/28$, which is
a 50 \% change from $1/41.6$, will make $\alpha_Z =1$ at
$\sim 2 \times 10^{2}\,eV$.)

For comparison, we also show two plots with $m_{\psi^{(Z)}_{2}} = 200\; GeV$
and $m_{\psi^{(Z)}_{1}}= 100\;GeV\,,\,50\;GeV$. From 
Fig. (\ref{fig7}, \ref{fig8}), we can see
that the change in the initial coupling $\alpha_Z(M)$ from the
previous case with $m_{\psi^{(Z)}_{2}} = 100\; GeV$ and
$m_{\psi^{(Z)}_{1}}= 50\;GeV$ is insignificant. This is shown in Table I.

Finally, we show in Table I and Fig. (\ref{fig9}) a result for
the $SU(2)_Z$ doublet messenger field. Here we observe that, for
the same range of masses, the initial coupling $\alpha_Z(M)$ is
approximately 11 \% smaller than the previous case. As we have
mentioned earlier, we will concentrate on the triplet case since
it is quite relevant to the SM leptogenesis proposed in our model.

It is interesting to note that the $SU(2)_Z$ coupling in various
cases shown in Figures (\ref{fig1}-\ref{fig9}) varies 
very little from its value at
the ``GUT'' scale $M$ to $E \sim m_{\psi^{(Z)}_{1,2}}$. In this
sense, the model is almost scale-invariant in the aforementioned
interval. This fact will be very useful in our discussion of
candidates in our model of the Cold Dark Matter.

In summary, we have seen in this section that, as can be seen from Table I
and Figs. (\ref{fig1}-\ref{fig8}), 
it is quite straightforward to have $SU(2)_Z$ strongly
interacting at a very low mass scale $\Lambda_Z = 3 \times 10^{-3}\,eV$.
Furthermore, for initial values of $\alpha_Z(M)$ close to the SM
couplings at comparable scales- which suggest some unification
with the SM at around that scale, the masses of $\psi^{(Z)}_{(L,R),1,2}$
are located in the region (e.g. $\sim 50-200\,GeV$) where the combination
of mass values as well as the strength of the $SU(2)_Z$ coupling (weak)
is such that they can become candidates for the WIMP cold dark matter,
a subject to which we will turn below. (We should keep in mind the
possibility that $\alpha_Z(M)$ can be anywhere in the range shown
in Figures (\ref{fig1}-\ref{fig8}), 
depending on the mass of $\psi^{(Z)}_{(L,R),1}$ and,
furthermore, on the pattern of the GUT breaking as mentioned below.)
But we will first discuss
the implication of the $SU(2)_Z$ scale $\Lambda_Z \sim 
3 \times 10^{-3}\,eV$ concerning the dark energy which is thought to be
responsible for the present accelerating universe.

We end this section by briefly mentioning the possibility
of unifying $SU(2)_Z$ with the SM. The most attractive route
in trying to achieve this unification is by noticing that
the famous GUT group $E_6$ contains $SU(2) \otimes SU(6)$. One
can envision the following symmetry breaking chain:
$E_6 \rightarrow SU(2)_Z \otimes SU(6) \rightarrow
SU(2)_Z \otimes SU(3)_c
\otimes SU(3)_L \otimes U(1) \rightarrow 
SU(3)_c \otimes SU(2)_L \otimes U(1)_Y
SU(2)_Z \otimes SU(3)_c \otimes U(1)_{em}$. 
A detailed study of
this scenario, including the symmetry breaking as well as the
evolution of the couplings, is in preparation \cite{hung2}.

\section{$SU(2)_Z$ and the Dark Energy}
\label{DE}

In this section, we will present a scenario in which the $SU(2)_Z$
axion is trapped in a false vacuum of an instanton-induced axion
potential and whose vacuum energy is $\sim (\Lambda_Z)^4$ \cite{su2}. 
We will then
present an estimate of the tunnelling probability to the true vacuum
and hence the lifetime of the false vacuum.
The basic assumption here is 
that we are currently
living in a {\em false vacuum} and that 
the associated vacuum energy relaxes to {\em zero}
once the phase transition is completed which will occur
in a very, very distant future according to our scenario
(see Eq. (\ref{time2})).

\subsection{The axion potential}
\label{axionpotential}

In Section (\ref{psizmass}), we present a discussion on the spontaneous
breakdown of the global $U(1)_{A}^{(Z)}$ symmetry of our model which is
responsible for giving masses to $\psi^{(Z)}_{1,2}$. This is due to
the non-vanishing VEV of a complex scalar field, 
$\phi_{Z} = v_{Z}\,\exp(ia_Z/v_{Z}) + \sigma_Z$, where 
$\langle \sigma_Z \rangle =0$ and
$\langle a_Z \rangle =0$. This results in a massive scalar, $\sigma_Z$,
and a massless Nambu-Goldstone (NG) boson, $a_Z$. (Notice the
following periodicity: $a_Z \rightarrow a_Z + 2\pi\,v_Z$.)
However, the $U(1)_{A}^{(Z)}$
global symmetry is explicitely broken by $SU(2)_Z$ instantons and
$a_Z$ becomes  a pseudo Nambu-Goldstone (PNG) boson with a
mass squared of order $\Lambda_{Z}^3/v_Z$
as we shall see below. This is quite similar to the famous Peccei-Quinn
axion. This axion, $a_Z$, is the ``quintessence'' field of our model. 

The instanton-induced axion potential has been calculated for the PQ axion
and can be straightforwardly applied to our model. At zero temperature,
one expects the axion potential, in the absence of a soft breaking term,
to look like $V(a_Z) \sim \Lambda_Z^4[1-\cos\frac{a_Z}{v_Z}]$ such that
$V(a_Z=0) =0$. However,
at temperatures $T \gg \Lambda_Z$, the $SU(2)_Z$ axion potential 
is flat because the contributions from $SU(2)_Z$ instantons and 
anti-instantons are suppressed \cite{instanton}. In fact, the instanton 
number density decreases drastically at high temperatures
as $n(\rho,T) \propto \exp(-[8\,\pi^2/g_Z^2 +
c\,(\pi\,\rho\,T)^{2}])$ where $c=2$ when $T >
m_{\psi^{(Z)}_{1,2}}$ and $c=4/3$ when
$T< m_{\psi^{(Z)}_{1,2}}$, with $\rho$ being the instanton size. 
Notice also the well-known factor $\exp(-8\,\pi^2/g_Z^2)$ 
which, for an asymptotically free theory like $SU(2)_Z$, increases 
as the temperature decreases since $g_Z^2$ does so.
One might parametrize this phenomenon
in the following way:
\be
\label{highT}
V(a_Z,T) = \Lambda_Z^4[1-\kappa(T)\,\cos\frac{a_Z}{v_Z}] \,,
\ee
where $\kappa(T)$ embodies the temperature dependence of the instanton
constribution. As we have mentioned earlier, we expect $\kappa(T)$
to rapidly decrease in magnitude as $T \gg \Lambda_Z$ and, as a result,
$V(a_Z,T \gg \Lambda_Z) \sim \Lambda_Z^4$. However, for $T \leq \Lambda_Z$,
one also expects $\kappa(T) \sim 1$ in which case $V(a_Z,T)$ exhibits 
two degenerate minima, one at $\langle a_Z \rangle =0$ and one
at $\langle a_Z \rangle = 2\pi v_Z$, with
the potential barrier between the two being $2\,\Lambda_Z^4$. 
This is due to the fact that
there is a remaining $Z(2)$ symmetry. Such degeneracy is well-known
in the PQ axion potential as it has been noted by \cite{sikivie}.
The computation of $\kappa(T)$
is fairly model-dependent. For our purposes, we only need to require
that $\kappa(T) \rightarrow 0$ for $T \gg \Lambda_Z$ and 
$\kappa(T) \sim 1$ for $T \leq \Lambda_Z$, noting that calculations
for the integrated instanton density at high temperatures as used
in the effective PQ axion potential show that it falls as $T^{-8}$.
In Figures (\ref{fig10},\ref{fig11}), we show $V(a_Z,T)$ 
for two values of $\kappa(T)$:
$\kappa(T)=1$ and $\kappa(T)=10^{-3}$ (as an illustrative value). 
Notice how quickly $V(a_Z,T)$
quickly flattens out at high temperature. (For that reason, we do not
show figures with $\kappa(T)< 10^{-3}$.)

In \cite{sikivie}, the $Z(N)$ degeneracy of the PQ axion potential
is lifted by a soft-breaking term (to evade the so-called domain wall
problem) of the form
$e^{i\delta}\,\mu^{3}\Phi + h.c.$, where $\Phi$ is a SM singlet
and $\mu^3 \ll \Lambda_{QCD}^4 / v_{\Phi}$. Similarly,
as in \cite{su2}, we would like to propose 
the following $U(1)_{A}^{(Z)}$ soft breaking term to lift
the $Z(2)$ degeneracy:
\be
\label{softbreaking}
V_B = \Lambda_Z^4\,\frac{a_Z}{2\pi\,v_Z}\,.
\ee
We shall assume a similar temperature dependence namely $\kappa(T)$
for $V_B$ such that for $T \gg \Lambda_Z$, the total effective
potential $V_{tot} = V(a_Z,T) + V_B(T)$ is flat and,
for $T \leq \Lambda_Z$ where we assume that $\kappa(T)=1$, it
is given by $V_{tot} = \Lambda_Z^4[1-\cos\frac{a_Z}{v_Z}] +
\Lambda_Z^4\,\frac{a_Z}{2\pi\,v_Z}$. We propose
\be
\label{totpot}
V_{tot}(a_Z,T) = \Lambda_Z^4[1-\kappa(T)\,\cos\frac{a_Z}{v_Z}] +
\kappa (T)\Lambda_Z^4\,\frac{a_Z}{2\pi\,v_Z} \,. 
\ee

\subsection{The false vacuum, its transition probability and
the equation of state}

We now discuss a cosmological scenario based on (\ref{totpot}).

1) We show in Figure (\ref{fig12}) $V_{tot}(a_Z,T)$ for $\kappa(T)=10^{-3}$. So,
at $T \gg \Lambda_Z$, the potential is flat. One might
expect the value of the classical $a_Z$ field to be 
$a_Z \sim O(v_Z)$. As long as the potential stays flat,
it will hover around that value as the temperature decreases.

2) We show in Figure (\ref{fig13}) $V_{tot}(a_Z,T)$ for $\kappa(T)=10^{-0.3}$.
We now see the appearance of two local minima: one
at $a_Z =0$ and the other (higher in energy) at 
$a_Z = 2\,\pi\, v_Z$. The latter is the false vacuum that we
had mentioned above. As $a_Z$ hovers around
$\sim O(v_Z)$ when the temperature decreases, it gets
trapped in the false vacuum when a local minimum develops at
$a_Z = 2\,\pi\, v_Z$.

3) In Figure (\ref{fig14}), we show $V_{tot}(a_Z,T)$ for $\kappa(T)=1$,
i.e. at $T \sim \Lambda_Z$. The true vacuum at $a_Z =0$
now has zero energy density and the barrier between the
two vacuua is now higher. The difference in
energy density between the true vacuum at $a_Z =0$ and the false
vacuum at $a_Z = 2\,\pi\, v_Z$ is $\Lambda_Z^4$. The universe
is still trapped in the false vacuum. How long does it stay there?

The first order phase transition to the true vacuum at $a_Z=0$
proceeds by bubble nucleation. The rate of the nucleation of the
true vacuum bubble is written as 
\be
\label{rate}
\Gamma = A \exp\{-S_E\}\,, 
\ee
where the Euclidean action $S_E$, in the thin wall limit, can be
computed by looking at 
\be
\label{Stil}
\tilde{S}=\int_{a_Z = 2 \pi v_Z}^{a_Z=0}
\sqrt{2(\Lambda_Z^4)[1-\cos\frac{a_Z}{v}]} da_Z= 8\,
v_Z\;\Lambda_{Z}^{2} 
\ee
giving
\be
\label{action}
S_E =  \frac{27\,\pi^2\,\tilde{S}^4}{2\,\Lambda_Z^{12}} \geq
5 \times 10^{5}\,(\frac{v_Z}{\Lambda_Z})^4 \,.
\ee
For $v_Z \sim $ a few hundreds of GeVs, the lower bound on $S_E$ is 
{\em huge}, approximately $10^{62}$!
The transition time can be estimated to be (with $T_c \sim \Lambda_Z$)
\be
\label{time}
\tau = \frac{3\,H}{4\,\pi \Gamma} \sim \{(\frac{\Lambda_Z}{m_{pl}})^2
\exp\{S_E\}\}\,t_{pl}  \,,
\ee
where we have taken $H \sim T_c^2/m_{pl}$ and $A \sim O(1)$. This gives
an estimate for the transition time to be approximately
\be
\label{time2}
\tau \geq (10^{-106}\,s) 
\exp(10^{62})\,.
\ee
A value of $\tau$ of this magnitude means that practically one is stuck in the
false vacuum for a very, very long time.

Assuming $a_Z$ to be spatially uniform, the equation of state parameter
$w$ is given by the well-known expression
\be
\label{w}
w(a_Z)= \frac{\frac{1}{2}\dot{a}_Z^2 - V(a_Z)}{\frac{1}{2}\dot{a}_Z^2 + V(a_Z)}\,.
\ee
When the universe is trapped in the false vacuum at $a_Z = 2 \pi v_Z$,
$\frac{1}{2}\dot{a}_Z^2 \sim 0$ and one obtains
\be
\label{w2}
w(a_Z) \approx -1 \,.
\ee
This means that the quintessence scenario presented here
{\em effectively mimics} the flat $\Lambda\,CDM$ model!

\subsection{Estimates of various ages of the universe in our scenario}
\label{age}

It is useful to estimate when the energy density ,$\Lambda_Z^4$, 
of the false vacuum started to dominate over the matter 
(baryonic and non-baryonic) energy density and when the deceleration
ceased and the acceleration kicked in. First, one
can readily estimate the temperature and time when $\rho_{vac} =\Lambda_Z^4$
equals the matter (baryonic and non-baryonic) energy density $\rho_M$
as follows. We start with the now accepted flat universe condition
\be
\label{omega}
\Omega_{M} + \Omega_{\Lambda_Z} =1\,,
\ee
where for definiteness we set the present values of $\Omega$'s to be
\be
\label{omega2}
\Omega_{M}^{0} = 0.3\;;\;\Omega_{\Lambda_Z}^{0}=0.7\,.
\ee
Since $\rho_M \propto T^3$ and $\rho_{vac} =\Lambda_Z^4$ is constant, it
follows that, with $T_{0} = 2.7^{0}\,K$,
\be
\label{temp}
T = T_{0}\,(\frac{0.7}{0.3})^{1/3}\,(\frac{\rho_M(T)}{\rho_{vac}})^{1/3}\,.
\ee
From (\ref{temp}), the temperature $T_{eq}$ at which $\rho_M(T_{eq})=\rho_{vac}$
is found to be
\be
\label{temp2}
T_{eq} \approx 3.6^{0}\,K\,.
\ee
In terms
of the redshift variable $z$, since $\rho_M \propto (1+z)^3$, one finds
\be
\label{z}
z_{eq} = (\frac{0.7}{0.3})^{1/3} - 1 \approx 0.33\,.
\ee
The age of the universe at a given redshift value $z$ is given by
\be
\label{timez}
t(z) = H_{0}^{-1} \int_{z}^{\inf} \frac{dz^{'}}{(1+z^{'})
[\Omega_{M}(1+z^{'})^{3} + \Omega_{vac}]^{1/2}}\,,
\ee
where $H_{0}^{-1} = (0.96 \pm 0.04)^{-1}\,t_0$ with $t_0 = 13 \pm 1.5
Gyr$ being the present age of the universe. Using Eq. (\ref{timez}), we
obtain the following age when the equality happened:
\be
\label{teq}
t_{eq} = 9.5 \pm 1.1 \,Gyr \,.
\ee

One may also want to know at what value of $z$ the deceleration ``stopped''
and the acceleration ``started''. With the equation for the cosmic
scale factor $a(t)$ being
\be
\label{at}
\frac{\ddot{a}}{a} = - \frac{4\pi\,G}{3} \sum_{i} \rho_{i}(1+3\,w_{i}) \,,
\ee
with $w_M=0$ and $w_{vac} = -1$,
the transition between the two regimes occured when $\ddot{a}=0$, giving
\be
\label{transition}
\rho_M(z_a)-2\,\rho_{vac}=0\,.
\ee
This gives
\be
\label{zt}
z_a = (\frac{2 \times 0.7}{0.3})^{1/3} - 1 \approx 0.67 \,.
\ee
The corresponding time and temperature are
\be
\label{ta}
t_a \approx 7.2 \pm 0.8\,Gyr\,,
\ee
\be
\label{Ta}
T_a \approx 4.5^{0}K\,.
\ee
The previous exercises serve two purposes: they give an estimate of the
time, temperature and redshift value of the period when the vacuum energy
density began to dominate over the matter energy density and the same
quantities for the period when the universe changed from a deceleration
stage to an accelerating one, and to compare the two. As one can see,
the acceleration began, in this scenario, about two billion years
before the dominance of the  vacuum energy density. As it is well-known,
both events occured rather ``recently''.

The next question is the temperature, time and redshift value of the epoch when
the axion potential developed a false local minimum, i.e.when 
$T_Z \sim \Lambda_Z$. One has to be however a little bit more careful 
here concerning the temperature of the $SU(2)_Z$ plasma as compared with
that of the SM plasma. 
At $T \gg m_i$, where $m_i$ is a generic particle mass, all
normal matter and matter that carries $SU(2)_Z$ quantum numbers
are in thermal equilibrium and are characterized by
a common temperature $T$. The fact that $SU(2)_Z$ matter is
in thermal equilibrium with normal matter is because the
messenger fields, 
${\tilde{\bm{\varphi}}}_{i}^{(Z)}$ or $\varphi^{(Z)}$, carry both SM and
$SU(2)_Z$ quantum numbers and, therefore, can interact
with normal matter as well as with the $SU(2)_Z$ ``gluons'' and
fermions $\psi^{(Z)}$. In what follows we will concentrate on
the scenario with ${\tilde{\bm{\varphi}}}_{i}^{(Z)}$ and the
estimates to be made below can be easily made for the $\varphi^{(Z)}$
case.

We first show that the energy density of the $SU(2)_Z$ plasma at
the time of Big Bang Nucleosynthesis (BBN) is a small fraction of
the SM plasma energy density. As a consequence, it does not affect BBN.
We then relate the temperatures of the two plasmas when the lightest
of the two $\psi^{(Z)}$'s drops out of thermal equilibrium.

When $T<m_{\tilde{\bm{\varphi}}^{(Z)}}$, ${\tilde{\bm{\varphi}}}_{1}^{(Z)}$
drops out of thermal equilibrium (how much might remain will be the subject
of the next section), $SU(2)_Z$ ``gluons'' and fermions $\psi^{(Z)}$
practically decouple from the SM plasma and its temperature $T_Z$ would
go like $R^{-1}$. To find its relationship with the SM temperature $T$,
we use the familiar expression for the effective number of relativistic
degrees of freedom $g_{\ast} = \sum g_{bosons} +(7/8) \sum g_{fermions}$.
Let us recall that $g_{\ast}$ for the SM above the top quark mass
is $g_{\ast}^{SM} = 427/4$ (right-handed neutrinos are not counted, being
SM singlets). To this we add the contibution from the $SU(2)_Z$ sector
when all particles are in thermal equilibrium. So for
$T>m_{\tilde{\bm{\varphi}}^{(Z)}}$, we obtain $g_{\ast}^{SU(2)_Z} = 39$
giving
\be
\label{gast}
g_{\ast}^{total}= \frac{583}{4}\,.
\ee
After ${\tilde{\bm{\varphi}}}_{1}^{(Z)}$ decouples, let us, for definiteness, 
label the temperature of the SM plasma by $T$ and that of the $SU(2)_Z$ plasma
by $T_Z$. The SM number of degrees of freedom for $m_e < T < m_{\mu}$ is
$g_{\ast}^{SM} = 43/4$. Furthermore, if the mass of the lightest
of the two $\psi^{(Z)}$'s, namely $\psi^{(Z)}_{1}$, is of O(GeV) as we
had discussed in Section (\ref{evolution}), the effective number of
$SU(2)_Z$ degrees of freedom after $\psi^{(Z)}_{1}$ decoupling is simply
$g_{\ast}^{SU(2)_Z}=6$, while it is $g_{\ast}^{SU(2)_Z}=27$ after 
${\tilde{\bm{\varphi}}}_{1}^{(Z)}$ decoupling. One obtains
\be
\label{temp3}
T_Z = (\frac{27}{6}\,\frac{43}{583})^{1/3}\,T \approx 0.7\,T\,.
\ee
This gives
\be
\label{rhoratio}
\frac{\rho_{SU(2)_Z}}{\rho_{SM}} = (6 \times \frac{4}{43})\times 
(\frac{27}{6}\,\frac{43}{583})^{4/3} \approx 0.13 \,.
\ee
From Eq. (\ref{rhoratio}), one can see that BBN is not affected by
the presence of the $SU(2)_Z$ plasma.

One can also relate $T_Z$ to the CMB temperature after $e^{\pm}$
decoupling, i.e. for $T < m_{e}$. (This is very similar to the relation between
the photon and neutrino temperatures.) It is given by
\be
\label{temp4}
T_Z = (\frac{27}{6}\,\frac{43}{583}\,\frac{4}{11})^{1/3}\,T
\approx 0.5\, T \,.
\ee
The $SU(2)_Z$ coupling grows strong ($\alpha_Z =1$) at $T_Z \sim
3 \times 10^{-3}\,eV \sim 35 ^{0}K$. Using Eq. (\ref{temp4}), one
can estimate the photon temperature at that point to be
\be
\label{temp5}
T \approx 70 ^{0}K\,.
\ee
This corresponds to a redshift
\be
\label{red5}
z \approx 25 \,.
\ee
The age of the universe at that point can be calculated using Eq. (\ref{timez})
to give
\be
\label{tz}
t_z \approx 125 \pm 14 Myr \,.
\ee

We now summarize the different epochs by listing the triplets of numbers:
redshift, age, and photon temperature. They are:

a) ($z \approx 25, t_z \approx 125 \pm 14 Myr, T \approx 70 ^{0}K$)
when $SU(2)_Z$ grew strong;

b) ($z_a \approx 0.67, t_a \approx 7.2 \pm 0.8 Gyr, T \approx 4.5 ^{0}K$)
when the acceleration ``kicked in'';

c) ($z_{eq} \approx 0.33, t_{eq} \approx 9.5. \pm 1.1 Gyr, T \approx 3.6 ^{0}K$)
when the energy density of the false vacuum equals that of (baryonic and non-baryonic)
matter.

From this summary, several remarks are in order.

1) According to our scenario, the universe got trapped in the false vacuum
of the $a_Z$ potential {\em long} before it began to accelerate. This
means that the mechanism which gives rise to the acceleration seven billion
years later occured at an age when the false vacuum energy density was
completely negligible compared with the matter energy density.

2) The fact that it started to accelerate six billion years ago (i.e. relatively
recent time) and that the dark energy density is comparable to
that of matter has to
do with the magnitude of the false vacuum energy density $\rho_{vac}
\sim (3 \times 10^{-3}\,eV)^4$. This is generic with any $\Lambda\,CDM$ model having
that value of vacuum energy density. Heuristically speaking, if a model
can generate a scale this low (i.e.at such a low temperature), the
acceleration process as well as the similarity in magnitudes of energy densities
necessarily took place fairly ``recently''.

3) This so-called coincidence (``why now'') problem is related to that particular
value of false vacuum energy density. In our scenario, this 
comes from the scale $\sim 10^{-3}\,eV$
where $SU(2)_Z$ grows strong as we have discussed in Section (\ref{evolution}).
In that section, the initial value of the $SU(2)_Z$ gauge coupling at some
``GUT'' scale of the order of $10^{16}\,GeV$ is taken to be comparable
to that of a SM coupling at a comparable scale. This is possible if
$SU(2)_Z$ were to merge with the SM into some grand unified group. There
is indeed one of such groups: the well-known $E_6$. One can envision
the following symmetry breaking pattern down to the SM: $E_6 \rightarrow
SU(2)_Z \otimes SU(6) \rightarrow SU(2)_Z \otimes SU(3)_c \otimes
SU(3)_L \otimes U(1)_1 \rightarrow SU(2)_Z \otimes SU(3)_c \otimes SU(2)_L
\otimes U(1)_2 \otimes U(1)_1 \rightarrow SU(2)_Z \otimes SU(3)_c \otimes SU(2)_L
\otimes U(1)_Y$. This path of $E_6$ breaking is very different from the usual
one where $E_6 \rightarrow SO(10)$. (The details of this scenario will
appear elsewhere \cite{hung2}.) From this pattern, one notices 
the following features:
(i) the $SU(2)_Z$ and $SU(6)$ couplings are equal at the scale of $E_6$
breaking which implies that the $SU(2)_Z$ coupling {\em is close to but
does not have to be equal} to the SM couplings at the same scale; (ii) the
unification of the SM itself follows a different route from the conventional
one where 3-2-1 merge into $SU(5)$ or $SO(10)$.

Based on the arguments presented above, one cannot help, within
the context of our model, but wonder if the dark energy might
be a direct remnant of the unification scenario discussed above. It is often
said that there appears to be an ``indirect evidence'' for Grand Unification
in the form of supersymmetric (SUSY) $SU(5)$ or $SO(10)$ by ``running''
the three SM couplings and finding that they all meet at one point
around $10^{16}\,GeV$. However, considering the possibility that
the path to unification can be more complicated than SUSY $SU(5)$ or $SO(10)$,
this ``indirect evidence'' based on the meeting of the three SM couplings
might be taken with caution. Perhaps the dark energy, or
equivalently the accelerating universe, could be the first
``direct evidence'' of Grand Unification?

\subsection{The mass of the quintessence axion field $a_Z$}

The last topic that we would like to discuss in this section is the
mass of the axion field $a_Z$. As we mentioned above, $a_Z$ would
be a massless NG boson if it were not for the fact that
the global $U(1)_{A}^{(Z)}$ symmetry  is explicitely broken by
$SU(2)_Z$ instantons. It then acquires a mass which can be computed
in a similar fashion to that for the PQ axion in QCD \cite{sikivie2}.

The mass of $a_Z$ can be computed by taking the vacuum expectation
value of the term $\sum_{i} K_{i} \, \bar{\psi}^{(Z)}_{L,i}\,\psi^{(Z)}_{R,i}\,\phi_{Z}
+ h.c.$ in Eq. (\ref{yuk}). From
\be
\label{axionvev}
\langle \sum_{i}| K_{i}| \, \bar{\psi}^{(Z)}_{L,i}\,\psi^{(Z)}_{R,i}\,\phi_{Z} + H.c. \rangle
= 2\,\sum_{i} |K_{i}|\, \mu_{i}^3\,v_{Z}\,\cos (\frac{a_Z}{v_Z}) \,,
\ee
and $\langle \bar{\psi}^{(Z)}_{L,i}\,\psi^{(Z)}_{R,i} \rangle = \mu_{i}^3$
one obtains the following mass squared for $a_Z$
\be
\label{axionmass}
m_{a_Z}^2 = \frac{2\,\sum_{i} |K_{i}|\, \mu_{i}^3}{v_Z} \,.
\ee
An approximate estimate of the upper bound of $m_{a_Z}$ can be found
by setting $\mu_{i}^3 \sim \Lambda_Z^3$ and $\sum_{i} |K_{i}| \alt 1$ 
in (\ref{axionmass}) giving
\be
\label{axionmass2}
m_{a_Z} \alt \Lambda_Z \, \sqrt{\frac{2\,\Lambda_Z}{v_Z}} \sim 10^{-10}\,eV \,,
\ee
where we have set $v_Z \sim 300\,GeV$ for simplicity. In fact, if we do not
want $K_{i}$ to be too much smaller than unity and since at least 
$m_{\psi^{(Z)}_{2}} \sim \mathcal{O}(100\,GeV)$, that choice is reasonable
for $v_Z$.

From Eq. (\ref{yuk}), one can write down the interaction term between $a_Z$
and the $SU(2)_Z$ fermions as follows
\be
\label{axionint}
{\cal L}_{a_Z} = i (\sum_{i} (\frac{m_{\psi^{(Z)}_{i}}}{v_Z})
\bar{\psi}^{(Z)}_{i}\,\gamma_{5} \psi^{(Z)}_{i})\,a_Z\,.
\ee
For $m_{a_Z} \sim 10^{-10}\,eV$, the range of interaction between two 
$\psi^{(Z)}$s is approximately $1\,km$. The astrophysical implication
of this interaction is under investigation.

We now turn our attention to two other cosmological implications of
our model: candidates for the Weakly Interacting Massive particles (WIMP) form 
of Cold Dark Matter (CDM), and a mechanism for leptogenesis. The latter
(leptogenesis) topic will appear as a companion paper while the former (CDM)
topic is under investigation. For this reason, the presentations which follow
will be brief.

\section{$\psi^{(Z)}_{1,2}$ as candidates for Cold Dark Matter}
\label{DM}

In this section, we will present a heuristic argument suggesting that
the $SU(2)_Z$ fermions $\psi^{(Z)}_{1,2}$ could be candidates for the 
cold dark matter, keeping in mind that the dark matter might very well
consist of a mixture of different particles.

As we have discussed above, at high temperatures $\psi^{(Z)}_{1,2}$ are
in thermal equilibrium with the $SU(2)_Z$ plasma as well as with the SM
plasma because of the presence of a messenger scalar field which,
for definiteness, we will take to be ${\tilde{\bm{\varphi}}}_{1}^{(Z)}$.

In principle, our scenario could contain several ``candidates'' for CDM:
${\tilde{\bm{\varphi}}}_{1}^{(Z)}$ and $\psi^{(Z)}_{1,2}$. As $T$
drops below various mass thresholds of these particles, they will
start the annihilation process that reduces their number. Out of
the three, ${\tilde{\bm{\varphi}}}_{1}^{(Z)}$ would be the most unstable
particle. In the next section, we will show that its decay which is
constrained to occur at a temperature $T_D$ larger that the electroweak
temperature $T_{EW} \sim 100\,GeV$ gives rise to a SM lepton
asymmetry which is reprocessed into a baryon asymmetry through
the electroweak sphaleron process. In what follows, we will only 
consider $\psi^{(Z)}_{1,2}$ as possible CDM candidates. 

Since we assume $m_{{\tilde{\bm{\varphi}}}_{1}^{(Z)}} >
m_{\psi^{(Z)}_{2}}> m_{\psi^{(Z)}_{1}}$, it is obvious that
$\psi^{(Z)}_{1}$ is {\em stable}. From (\ref{yuk}), one can see that
$\psi^{(Z)}_{2}$ can decay into $\psi^{(Z)}_{1} + {\bf A}^{(Z)}$
via a one-loop diagram. A rough estimate of the decay rate of
$\psi^{(Z)}_{2}$ gives $\Gamma_{\psi^{(Z)}_{2}} \sim 
(\alpha_{{\tilde{\varphi}_{1}}^{1}}
\alpha_{{\tilde{\varphi}_{1}}^{2}}\,\alpha_Z)\,m_{\psi^{(Z)}_{2}}$
which could be less than the Hubble rate if the Yukawa couplings in (\ref{yuk})
are small enough. 
Furthermore, if the Yukawa couplings are sufficiently
small so that the interactions freeze out before it decays, 
$\psi^{(Z)}_{2}$ could be considered to be stable. When $T<m_{\psi^{(Z)}_{2}}$,
the number density of $\psi^{(Z)}_{2}$ decreases like $exp(-m/T)$ until
its annihilation rate drops below the Hubble rate and $\psi^{(Z)}_{2}$
drops out of thermal equilibrium. To find out about the relic abundance
of $\psi^{(Z)}_{2}$ and of $\psi^{(Z)}_{1}$, one needs to know 
the size of the annihilation cross sections. 

One of the most attractive candidates for CDM is a stable
particle which has an annihilation cross section typically the size
of the electroweak cross section. Although a detailed analysis is needed
in order to make a more precise prediction, an insight can be gained
in this section by noticing that an approximate solution to the
Boltzmann equation gives the following estimate for the fraction
of the energy density coming from the relic abundance \cite{kamionkowski}
\be
\label{omegapsi2}
\Omega_{\psi^{(Z)}_{2}} = \frac{m_{\psi^{(Z)}_{2}}\,n_{\psi^{(Z)}_{2}}}
{\rho_{c}\,h^2} \approx \bigl(\frac{3 \times 10^{-27}\,cm^{3}\,sec^{-1}}
{\langle \sigma_{A,\psi^{(Z)}_{2}}\,v\rangle} \bigr) \,, 
\ee
where $\rho_c = 3\,H^2/8\,\pi\,G$ is the critical density,
$h \approx 0.72$ and $\sigma_{A,\psi^{(Z)}_{2}}$ is the
annihilation cross section for $\sigma_{A,\psi^{(Z)}_{2}}$. 
In this approximation, (\ref{omegapsi2}) is
{\em independent} of the $\psi^{(Z)}_{2}$ mass and depends only on its
annihilation cross section \cite{kamionkowski}. For this reason
and using the same approximation, we infer that
\be
\label{omegapsi3}
\Omega_{\psi^{(Z)}_{1}} = \frac{m_{\psi^{(Z)}_{1}}\,n_{\psi^{(Z)}_{1}}}
{\rho_{c}\,h^2} \approx \bigl(\frac{3 \times 10^{-27}\,cm^{3}\,sec^{-1}}
{\langle \sigma_{A,\psi^{(Z)}_{1}}\,v\rangle} \bigr) \,. 
\ee

In order for $\Omega_{\psi^{(Z)}_{2}}$ and/or $\Omega_{\psi^{(Z)}_{1}}$ or
$\Omega_{\psi^{(Z)}_{2}}+\Omega_{\psi^{(Z)}_{1}}$ to be of order unity, the
annihilation cross sections $\sigma_{A,\psi^{(Z)}_{1,2}}$  should have
a magnitude of the order $3 \times 10^{-27}\,cm^{3}\,sec^{-1}/\langle v \rangle$.
Although $\psi^{(Z)}_{1,2}$ are non-relativistic
when $T$ drops below their masses, $\langle v \rangle$ might not be too small.
For the sake of estimate, let us assume $\langle v \rangle \sim 0.1\,c$.
A typical magnitude for the annihilation cross sections so that the relic
abundance of CDM is of the right order would be
\be
\label{xsection1}
\langle \sigma_{A,\psi^{(Z)}_{1,2}} \rangle \sim 10^{-36}\, cm^2 \sim \frac{3 \times
10^{-9}}{GeV^2}\,.
\ee
Under what conditions would $\sigma_{A,\psi^{(Z)}_{1,2}}$ have
a magnitude $\sim 10^{-36}\, cm^2 \sim \frac{3 \times
10^{-9}}{GeV^2}$? In our model, the dominant annihilation
cross section goes like
\be
\label{xsection2}
\sigma_{A,\psi^{(Z)}} \sim \frac{\alpha_{Z}(T)^2}{T^2} \,,
\ee
for $T<m_{\psi^{(Z)}}$. 
From Fig. (\ref{fig1},
\ref{fig2}, \ref{fig3}, \ref{fig4}, \ref{fig5}), we can see that 
$\alpha_{Z}(T)$ is relatively ``flat''  
for a large range of energy, ranging from $\sim 10^{16}\,GeV$ down to
approximately $m_{\psi^{(Z)}_{2}}$ where it begins to ``rapidly'' increase.
We also see that this ``flat'' value for $\alpha_{Z}(T)$ depends on
the mass of $\psi^{(Z)}_{1}$ for $m_{\psi^{(Z)}_{2}} \sim \mathcal{O}(100\,GeV)$.
Typically, $\alpha_{Z}(T) \sim 1/30-1/40$ for $m_{\psi^{(Z)}_{1}} \sim 1\,eV-
50\,GeV$ according to Fig. (\ref{fig1},
\ref{fig2}, \ref{fig3}, \ref{fig4}, \ref{fig5}).
We can now make two generic remarks.

1) From (\ref{xsection2}) and from $\alpha_{Z}(T)^2 \sim 6 \times 10^{-4}$,
one can infer that $T<m_{\psi^{(Z)}} <1\,TeV$ otherwise $\psi^{(Z)}$ will
become overabundant. 

2) $\psi^{(Z)}_{1}$ cannot be too light e.g. $< 10 \,GeV$ or so since this would
lead to a cross section which could be too large and which could greatly reduce
its relic abundance. Therefore,if it
were to be a CDM candidate, its mass should be high enough in value in order
for the cross section to be of the right order of magnitude since $\alpha_{Z}(T)^2$
is practically ``constant'' in the interval of interest.

The above discussion makes clear that, whether or not $\psi^{(Z)}_{1,2}$ 
can be considered to be reasonable WIMP CDMs, 
it is a question which actually depends on the masses of these particles 
through the magnitude of their
annihilation cross sections. Furthermore, the best range of masses appears
to be of $\mathcal{O}(100-1000\,GeV)$. It is interesting to note the following
fact. As we have seen in Section (\ref{evolution}),
this range of masses for {\em both} $\psi^{(Z)}_{1}$ and $\psi^{(Z)}_{2}$ gives
a value for the initial $SU(2)_Z$ coupling $\alpha_Z(M)$ $\sim 1/42$ which
is very close to the (non-supersymmetric) SM couplings at a similar scale, 
which suggests some kind of
unification as we had mentioned earlier. (A full investigation of the unification
issue is slightly more complicated.)
The next question is the following: Which
of the $\psi^{(Z)}$s is the best candidate or is it both? Below we list two
possible scenarios with one being more attractive than the other.

\bi

\item $m_{\psi^{(Z)}_{2}} \sim \mathcal{O}(100\,GeV)$, $m_{\psi^{(Z)}_{1}}
\sim \mathcal{O}(<10\,GeV)$: 

Here it is unlikely for $\psi^{(Z)}_{1}$ to be a WIMP because of its
mass but $\psi^{(Z)}_{2}$ could. However, as we have noted above, 
$\psi^{(Z)}_{2}$ decays into $\psi^{(Z)}_{1} + {\bf A}^{(Z)}$. The
decay rate which arises at one loop depends very much on the strength
of the Yukawa couplings in (\ref{yuk}), in particular the one
involving $\psi^{(Z)}_{1}$ . The decay rate of $\psi^{(Z)}_{2}$ is
approximately $\Gamma_{\psi^{(Z)}_{2}} \sim (\alpha_{Z}\,
\alpha_{{\tilde{\bm{\varphi}}}_{1}^{(Z)}}\,\alpha_{{\tilde{\bm{\varphi}}}_{2}^{(Z)}})
(m_{\psi^{(Z)}_{2}}/m_{{\tilde{\bm{\varphi}}}_{1}^{(Z)}})^{4}
(m_{\psi^{(Z)}_{2}}/16\,\pi^2)$. For $m_{\psi^{(Z)}_{2}}\sim \mathcal{O}
(m_{{\tilde{\bm{\varphi}}}_{1}^{(Z)}})$, one notices that $\psi^{(Z)}_{2}$
can only survives until the present time ($\sim 4.3 \times 10^{17}\,sec$ if
$\alpha_{{\tilde{\bm{\varphi}}}_{1}^{(Z)}}\,\alpha_{{\tilde{\bm{\varphi}}}_{2}^{(Z)}}
\sim 10^{-41}$ which might be quite unnatural. Otherwise $\psi^{(Z)}_{2}$
will decay out-of-equilibrium at some earlier times. Whether or not the
decay process preserves the desired fraction of the total energy density
is beyond the scope of this paper and will be presented elsewhere.

\item $m_{\psi^{(Z)}_{2}} \sim m_{\psi^{(Z)}_{1}} \sim \mathcal{O}(100\,GeV)$,
e.g. $m_{\psi^{(Z)}_{2}}= 200\,GeV$ and $m_{\psi^{(Z)}_{1}}= 100\,GeV$ as
shown in Fig. (\ref{fig7}):

This case appears to be the more desirable one.
The lighter of the two and hence the stable one, namely $\psi^{(Z)}_{1}$, 
has a mass of $\mathcal{O}(100\,GeV)$ and, from the results of the above discussion,
can have the desired relic abundance. Alternatively, a combination of $\psi^{(Z)}_{1}$
and $\psi^{(Z)}_{2}$ (or its decay product) can have the desired abundance. Since
$\psi^{(Z)}_{2}$ decays, the principal WIMP candidate is actually $\psi^{(Z)}_{1}$.

\ei

One last remark we would like to make in this section concerns the present form of
the WIMP candidate(s) of our model. As we discuss above, $SU(2)_Z$ grows strong
at $\Lambda_Z \sim 10^{-3}\,eV$. If we assume that this leads to confinement
as with QCD, the $SU(2)_Z$ singlets would be a {\em spin zero} composite of two
$\psi^{(Z)}$s. In priciple, one would also have a spin one-half composite
of one $\psi^{(Z)}$ and a messenger field. However, the messenger field decays
and practically disappears long before this ``confinement'' occurs. Therefore,
the present form of WIMP in our model would be a chargeless, 
spin zero $SU(2)_Z$ ``hadron'' whose phenomenological implication is
briefly discussed in Section (\ref{phenomenology}). Presumably
the size of this ``hadron'' would be of the order $\hbar c/10^{-3}\,eV
\sim 1\,mm$ which is rather large.

\section{$SU(2)_Z$ and leptogenesis}
\label{lepto}

In this section, we will present a brief discussion of the possibility of
leptogenesis in our model. A full presentation will appear in a companion
article.

As we have presented above, our model contains two $SU(2)_Z$ triplet
complex scalar fields, $\tilde{\bm{\varphi}}_{1}^{(Z)}$ and
$\tilde{\bm{\varphi}}_{2}^{(Z)}$. These fields interact with
$\psi^{(Z)}_{1,2}$ and the SM leptons via Eq. (\ref{yuk}). In
Section (\ref{evolution}), one of the two messenger fields,
$\tilde{\bm{\varphi}}_{2}^{(Z)}$, was set to have a large mass of
the order of the ``GUT'' scale and the evolution of the $SU(2)_Z$
coupling on the messenger fields depends only on 
$\tilde{\bm{\varphi}}_{1}^{(Z)}$
as well as on the $SU(2)_Z$ fermions. $\tilde{\bm{\varphi}}_{1}^{(Z)}$
can decay into $\psi^{(Z)}_{1,2}$ plus a SM lepton. The
interference between the tree-level and one-loop amplitude for the
previous decay generates a SM lepton number violation which transmogrifies
into a baryon asymmetry through the electroweak sphaleron process
\cite{kuzmin}, \cite{lepto}.
An important point to keep in mind is the fact that $\psi^{(Z)}_{1,2}$
are SM singlets and the $\psi^{(Z)}_{1,2}$ number violation cannot
be reprocessed by the electroweak sphaleron. So the rule of thumb
here is the following: SM lepton number violation $\rightarrow$
quark (or baryon) number violation.

$\tilde{\bm{\varphi}}_{1}^{(Z)}$ is in thermal equilibrium (with
the $SU(2)_Z$ as well as with the SM plasmas) at $T>
m_{{\tilde{\bm{\varphi}}}_{1}^{(Z)}}$. When $T \approx
m_{{\tilde{\bm{\varphi}}}_{1}^{(Z)}}$, one would like 
$\tilde{\bm{\varphi}}_{1}^{(Z)}$ to decouple before it decays.
The primary condition for a departure from thermal equilibrium is the 
requirement that the decay rate $\Gamma_{{\tilde{\bm{\varphi}}}_{1}^{(Z)}}
\sim \alpha_{\tilde{\varphi}_{1}} m_{\tilde{\varphi}_{1}}$,
with $\alpha_{\tilde{\varphi}_{1}} = g_{\alpha_{\tilde{\varphi}_{1}}}^2/4\pi$, 
is {\em less than} the expansion rate $H = 1.66\, g_{*}^{1/2} T^2/m_{pl}$,
where $g_{*}$ is the effective number of degrees of freedom at 
temperature $T$. As with \cite{kolb}, we can define
\be
\label{K}
K \equiv (\Gamma_{\tilde{\varphi}_{1}}/2\,H)_{T=m_{\tilde{\varphi}_{1}}}
= \frac{\alpha_{\tilde{\varphi}_{1}}\,m_{pl}}{3.3\,g_{*}^{1/2}\,
m_{{\tilde{\bm{\varphi}}}_{1}^{(Z)}}} \,.
\ee
When $K \ll 1$, $\tilde{\bm{\varphi}}_{1}^{(Z)}$ and 
$\tilde{\bm{\varphi}}_{1}^{(Z),*}$ are
overabundant and depart from thermal equilibrium. Since the time
when $\tilde{\bm{\varphi}}_{1}^{(Z)}$ decays is $t \sim 
\Gamma^{-1}_{\tilde{\bm{\varphi}}_{1}^{(Z)}}$ 
and since $T \propto 1/\sqrt{t}$, the
temperature at the time of decay is found to be (using (\ref{K}))
$T_D \sim K^{1/2}\,m_{\tilde{\bm{\varphi}}_{1}^{(Z)}}$ \cite{kolb}. For this
scenario to be effective i.e. a conversion of a SM lepton number asymmetry
coming from the decay of $\tilde{\bm{\varphi}}_{1}^{(Z)}$ into a baryon number
asymmetry through the electroweak sphaleron process, one has to make sure
that the decay occurs at a temperature greater than $T_{EW} \sim
100\,GeV$ above which the sphaleron processes are in thermal equilibrium.
From this, it follows that $K$ cannot be arbitrarily small and
has a lower bound coming from the requirement $T_D > T_{EW}$. One obtains
\be
\label{lower}
1>\,K\, > (\frac{100\,GeV}{m_{\tilde{\bm{\varphi}}_{1}^{(Z)}}})^2 \,.
\ee
This translates into $1>K \agt 0.1$, with the lower bound getting
smaller as we increase $m_{\tilde{\bm{\varphi}}_{1}^{(Z)}}$.

When $T<m_{\tilde{\varphi}_{1}}$ and when $K<1$,
the number density of $\tilde{\varphi}_{1}$ is approximately
$n_{\tilde{\varphi}_{1}} =T^{3}/\pi^2$ (overabundance) and the entropy is
$s = (2/45)\,g_{*} \pi^2\,T^3$, with $g_{*} \sim 114$
(including $SU(2)_Z$ light degrees of freedom). 
The decay of $\tilde{\varphi}_{1}$ and $\tilde{\varphi}_{1}^{*}$
creates a SM lepton number asymmetry per unit entropy 
$n_{LSM}/s \sim 2 \times 10^{-3}\,\epsilon^{\tilde{\varphi}_{1}}_{l}$.
For the SM with three generations and one Higgs doublet, one
has $n_{B} \sim -0.35\,n_{LSM}\,\sim -10^{-3}\,\epsilon^{\tilde{\varphi}_{1}}_{l}$,
where $n_B$ is ``processed'' through the electroweak sphaleron.
Since $m_{B}/s \sim 10^{-10}$, a rough constraint on 
$\epsilon^{\tilde{\varphi}_{1}}_{l}$ is found to be
\be
\label{constraint}
\epsilon^{\tilde{\varphi}_{1}}_{l} \sim -10^{-7} \,.
\ee

One can now calculate $\epsilon^{\tilde{\varphi}_{1}}_{l}$ and use
the constraint (\ref{constraint}) to restrict the range of parameters
involved in the calculation which is carried out in (). In that
companion article, $\epsilon^{\tilde{\varphi}_{1}}_{l}$ is calculated
at $T=0$. Although, care should be taken to include finite temperature
corrections (see e.g. \cite{giudice}), one expects the final
result not to be too different from the zero temperature one.
$\epsilon^{\tilde{\varphi}_{1}}_{l}$ is defined as
\be
\label{asymmetry}
\epsilon^{\tilde{\varphi}_{1}}_{l} = \frac{\Gamma_{\tilde{\varphi}_{1}\,l}-
\Gamma_{\tilde{\varphi}_{1}^{*}\,\bar{l}}}{\Gamma_{\tilde{\varphi}_{1}\,l}+
\Gamma_{\tilde{\varphi}_{1}^{*}\,\bar{l}}} \,,
\ee
where $\Gamma_{\tilde{\varphi}_{1}\,l}$ and $\Gamma_{\tilde{\varphi}_{1}^{*}\,
\bar{l}}$ contain the sums over all three flavors of SM leptons. A non-vanishing
value for $\epsilon^{\tilde{\varphi}_{1}}_{l}$ in (\ref{asymmetry}) in
the interference between the tree-level and one-loop diagrams. The details
of the calculations are presented in (\cite{hung3}). 
We will present here a brief summary
of some salient features of the results that are obtained there. It turns
out that the dominant contribution to $\epsilon^{\tilde{\varphi}_{1}}_{l}$ takes
approximately the following form (a full expression can be found in (\cite{hung3})):
\be
\label{asymmetry2}
\epsilon^{\tilde{\varphi}_{1}}_{l} \approx \sum _{i}\,f(g_{\tilde{\varphi}_{1}\,i},
g_{\tilde{\varphi}_{2}\,i},\theta_{i})\,Im\{\delta\,I_{i}\}\,,
\ee
where the function $f$ on the right-hand side of (\ref{asymmetry2}) contains
the dependence on the various Yukawa couplings and phases 
and is given in (\cite{hung3}),
$\theta_i$ are the phase angles and $i=e,\mu,\tau$. The function 
$Im\{\delta\,I_{i}\}$ is $\sim -\frac{1}{8\pi}(\frac{m_{l_i}}
{m_{\tilde{\varphi}_{1}}})
(\frac{m_{\psi^{(Z)}_{1,2}}}{m_{\tilde{\varphi}_{1}}})^{3}$ with
$Im\{\delta\,I_{\tau}\}$ being the dominant one. In many cases which are
examined in (\cite{hung3}), $\epsilon^{\tilde{\varphi}_{1}}_{l}$ is found to
depend pricipally on the Yukawa couplings $g_{\tilde{\varphi}_{2}\,i}$ between
the {\em heavier} of the two messenger fields, $\tilde{\bm{\varphi}}_{2}^{(Z)}$,
to the $SU(2)_Z$ fermions and the SM leptons. Using (\ref{constraint})
and various general arguments, we concluded in (\cite{hung3}) that the mass of the
decaying and lighter messenger field, $\tilde{\bm{\varphi}}_{1}^{(Z)}$, is
bounded from above by approximately $700\,GeV-1\,TeV$. This upper
bound on the $\tilde{\bm{\varphi}}_{1}^{(Z)}$ mass in conjunction
with the values used in the evolution of the $SU(2)_Z$ coupling,
namely $m_{\tilde{\bm{\varphi}}_{1}^{(Z)}} \sim \mathcal{O}(300\,GeV-1\,TeV)$,
makes it possible to search for signals of the light messenger field at
future colliders. We will briefly discuss these phenomenological
issues below.

\section{Other phenomenological consequences of $SU(2)_Z$}
\label{phenomenology}

In addition to providing a model for the dark energy and dark matter
as well as a mechanism for SM leptogenesis, one might ask
whether or not one can detect any of the $SU(2)_Z$ particles
in earthbound laboratories, namely $\tilde{\bm{\varphi}}_{1}^{(Z)}$
as well as $\psi^{(Z)}_{1,2}$. 
Let us recall that, under $SU(2)_L
\otimes SU(2)_Z$, these particles transform as
$\tilde{\bm{\varphi}}_{1}^{(Z)}= (2,3)$ and $\psi^{(Z)}_{1,2} = (1,3)$.
Therefore, only $\tilde{\bm{\varphi}}_{1}^{(Z)}$ can be produced
at tree level by the electroweak gauge bosons. $\psi^{(Z)}_{1,2}$,
being electroweak singlets, can only interact with the SM matter
either through (\ref{yuk}) or through its magnetic moment.

In the kinetic terms for the messenger fields,
and in particular for $\tilde{\bm{\varphi}}_{1}^{(Z)}$, one is interested in the
following interactions: $W^{+}\,W^{-}\,(\tilde{\bm{\varphi}}_{1}^{(Z),0*}
\tilde{\bm{\varphi}}_{1}^{(Z),0}+\tilde{\bm{\varphi}}_{1}^{(Z),+}\,
\tilde{\bm{\varphi}}_{1}^{(Z),-})$
and $Z\,Z\,(\tilde{\bm{\varphi}}_{1}^{(Z),0*}
\tilde{\bm{\varphi}}_{1}^{(Z),0}+\tilde{\bm{\varphi}}_{1}^{(Z),+}\,
\tilde{\bm{\varphi}}_{1}^{(Z),-})$.
These interactions will provide the dominant weak boson fusion (WBF)
production mechanism for a pair of $\tilde{\bm{\varphi}}_{1}^{(Z)}$. A rough
expectation for the production cross section for $\tilde{\bm{\varphi}}_{1}^{(Z)}$
with a mass around $300\,GeV$ is around $1\,pb$. The decay
$\tilde{\bm{\varphi}}_{1}^{(Z),0} \rightarrow \bar{\psi}^{(Z)}_{2}+ l^{0}_{i}$
is practically unobservable while
$\tilde{\bm{\varphi}}_{1}^{(Z),-} \rightarrow \bar{\psi}^{(Z)}_{2}+ l^{-}_{i}$
and $\tilde{\bm{\varphi}}_{1}^{(Z),+} \rightarrow \psi^{(Z)}_{2}+ l^{+}_{i}$
will have charged SM leptons with unconventional geometry, perfectly
distinguishable from the decay of a $600\,GeV$ SM Higgs boson. It is
also useful to estimate the length of the charged tracks left
by $\tilde{\bm{\varphi}}_{1}^{(Z),\pm}$ before they decay. We will
focus on $m_{\tilde{\bm{\varphi}}_{1}^{(Z)}} = 300\, GeV$ which is used
as an example in this paper, leaving other values to a more detailed
phenomenological analysis which will appear elsewhere. As we have
discussed above, the lifetime of $\tilde{\bm{\varphi}}_{1}^{(Z)}$ is
constrained by the quantity $K$ defined in (\ref{K},\ref{lower}).
The constraint (\ref{lower}) gives 
$10^{-16} \alt \alpha_{\tilde{\varphi}_{1}} \alt 2 \times 10^{-15}$.
Since $\Gamma_{{\tilde{\bm{\varphi}}}_{1}^{(Z)}}
\sim \alpha_{\tilde{\varphi}_{1}} m_{\tilde{\varphi}_{1}}$, the
decay lengths are approximately $0.02\,cm\alt l_{\tilde{\varphi}_{1}}\alt 1\,cm$,
which are within the range of the radial
region of a typical silicon detector at CMS 
and ATLAS ($40\,cm$ and $60\,cm$ respectively).
 
As we discussed above, $\psi^{(Z)}_{1,2}$ could be WIMP CDMs and
their detection falls into the domain of dark matter search. A
study is in progress concerning various direct signals such
as: $\psi^{(Z)}_{1,2} + e \rightarrow e + \psi^{(Z)}_{1,2}$, where
$e$ is an atomic electron (e.g. in a Rydberg atom);
$\psi^{(Z)}_{1,2} + N \rightarrow \psi^{(Z)}_{1,2} + N$, where $N$
is a nucleon, which can occur through the magnetic moment of
$\psi^{(Z)}_{1,2}$. Also under investigation is the possibility of
$\mu - e$ conversion in our model involving the interaction of 
muons with nuclei, a process which can occur at the one-loop level.
As we mentioned in Section (\ref{DM}), the present form of
our WIMP candidate would be a chargeless, 
spin zero $SU(2)_Z$ ``hadron'' which is a composite of
two $\psi^{(Z)}$s and which is of a milimeter size.

The above discussion represents only a few of several phenomenological
implications of the $SU(2)_Z$ model which could be tested in future
accelerators and dedicated detectors. 

\section{Conclusion}

We have presented a model involving a new unbroken gauge group $SU(2)_Z$
which becomes strongly interacting at a scale $\Lambda_Z \sim 10^{-3}\,eV$,
starting with a value for the gauge coupling, at a high scale $\sim 10^{16}
\,GeV$, which is close to that of a typical SM coupling at a similar scale.
This similarity in gauge couplings at high energies is suggestive of
a unification between $SU(2)_Z$ and the SM. A possible scenario
for such a unification is briefly discussed here. 

There are several cosmological implications of the $SU(2)_Z$ model.
The most important one is a quintessence model for dark energy in
which the quintessence field is the Peccei-Quinn-like axion $a_Z$
whose potential is induced by the $SU(2)_Z$ instantons. Unlike 
other quintessence models, our scenario involves the existence
of a false vacuum where the $SU(2)_Z$ axion is trapped as the 
$SU(2)_Z$ plasma is cooled to the temperature $T \sim \Lambda_Z$.
This occured when the age of the universe is
$t_z \approx 125 \pm 14 Myr$ (at redshift $z \sim 25$). The
age when the acceleration began was computed to be
$t_a \approx 7.2 \pm 0.8\,Gyr$ (redshift $z \sim 0.67$). The energy density of
the false vacuum started to dominate the (baryonic and non-baryonic)
matter density at around $t_{eq} = 9.5 \pm 1.1 \,Gyr$ (redshift
$z \sim 0.33$). Since the universe is trapped in the false vacuum,
the equation of state $w(a_Z) \approx -1$. 
This means that the quintessence scenario presented here
{\em effectively mimics} the flat $\Lambda\,CDM$ model! The
most recent supernovae results (up to redshift $z=1$) when combined
with those from the Sloan Digital Sky Survey fits a flat
$\Lambda\,CDM$ model with $w \approx -1$.

There are two other cosmological consequences of our model: 
1) The $SU(2)_Z$ fermions $\psi^{(Z)}_{1,2}$ as candidates
of Weakly Interacting Massive Particles (WIMP) cold dark
matter; 2) The decay of the messenger scalar field
$\tilde{\bm{\varphi}}_{1}^{(Z)}$ into $\psi^{(Z)}_{1,2}$ plus
a SM lepton generating a SM lepton asymmetry which transmogrifies
into a baryon asymmetry through the electroweak sphaleron process.
For (1), we showed that, with the masses of $\psi^{(Z)}_{1,2}$
of $\mathcal{O}(100\,GeV)$, not only one obtains the initial
(high energy) value of the $SU(2)_Z$ gauge coupling to be
close in value to those of the SM couplings at a similar scale, one
also finds that, when the temperature drops below their masses, 
the annihilation cross section is typically of the size of a weak
cross section which is what is usually required in order for
the relic abundances of these particles to be of the order
of the ``observed'' CDM abundance. For (2), we showed that
the interference between the tree-level and one-loop decay
rates of $\tilde{\bm{\varphi}}_{1}^{(Z)}$ into $\psi^{(Z)}_{1,2}$ plus
a SM lepton gives rise to a non-vanishing SM lepton asymmetry,
which can be subsequently transformed into a baryon asymmetry. We
then showed that, in order for this to happen, $\tilde{\bm{\varphi}}_{1}^{(Z)}$
has to be lighter than $\sim 1\,TeV$. Since
$\tilde{\bm{\varphi}}_{1}^{(Z)} = (2,3)$ under $SU(2)_L \otimes SU(2)_Z$,
this mass constraint opens up the possibility of detecting the
messenger fields at the LHC (or other future colliders). The
details of the leptogenesis scenario are presented in a companion article
\cite{hung3}. 

Finally, we end the paper with a brief discussion of the detectability
of the messenger scalar field as well as other processes involving
the CDM candidates $\psi^{(Z)}_{1,2}$. In particular, we showed that
the production and subsequent decay of the messenger field shows
characteristic signals in terms of the decay geometry as well as
the length of the charged tracks. The possible detection
of $\psi^{(Z)}_{1,2}$ as CDM matter as well as its contribution
to a process such as $\mu-e$ conversion present interesting
phenomenological challenges which are under investigation.

Note added: After this present paper was completed, I learned from 
James ({\em bj}) Bjorken that an earlier paper by Larry Abbott 
\cite{abbott} contained some ideas
which are similar in spirit to those presented here. It would be
interesting to see if one can apply our model to
the idea of a ``compensating field'' presented in \cite{abbott}.

\begin{acknowledgments}

I would like to thank
{\em bj} for bringing my attention to \cite{abbott}.
I also wish to thank Lia Pancheri, Gino Isidori and
the Spring Institute for the hospitality in the
Theory Group at LNF, Frascati, where part of this work
was carried out.
This work is supported in parts by the US Department
of Energy under grant No. DE-A505-89ER40518. 
\end{acknowledgments}

\begin{figure}
\includegraphics[angle=-90,width=8cm]{alvst_50.epsi}
\includegraphics[angle=-90,width=8cm]{alinvst_50.epsi} 
\caption{\label{fig1}$\alpha_Z(E)$ and $\alpha_Z^{-1}(E)$ versus
$t=\ln (E/\Lambda_Z)$ for $m_{\tilde{\bm{\varphi}}_{1}^{(Z)}}
= 300\,GeV$, $m_{\psi^{(Z)}_{2}} = 100\; GeV$ and
$m_{\psi^{(Z)}_{1}}= 50\;GeV$.}
\end{figure}
\begin{figure}
\includegraphics[angle=-90,width=8cm]{alvst_10.epsi}
\includegraphics[angle=-90,width=8cm]{alinvst_10.epsi} 
\caption{\label{fig2}$\alpha_Z(E)$ and $\alpha_Z^{-1}(E)$ versus
$t=\ln (E/\Lambda_Z)$ for $m_{\tilde{\bm{\varphi}}_{1}^{(Z)}}
= 300\,GeV$, $m_{\psi^{(Z)}_{2}} = 100\; GeV$ and
$m_{\psi^{(Z)}_{1}}= 10\;GeV$.}
\end{figure}
\begin{figure}
\includegraphics[angle=-90,width=8cm]{alvst_1.epsi}
\includegraphics[angle=-90,width=8cm]{alinvst_1.epsi} 
\caption{\label{fig3}$\alpha_Z(E)$ and $\alpha_Z^{-1}(E)$ versus
$t=\ln (E/\Lambda_Z)$ for $m_{\tilde{\bm{\varphi}}_{1}^{(Z)}}
= 300\,GeV$, $m_{\psi^{(Z)}_{2}} = 100\; GeV$ and
$m_{\psi^{(Z)}_{1}}= 1\;GeV$.}
\end{figure}
\begin{figure}
\includegraphics[angle=-90,width=8cm]{alvst_1MEV.epsi}
\includegraphics[angle=-90,width=8cm]{alinvst_1MEV.epsi} 
\caption{\label{fig4}$\alpha_Z(E)$ and $\alpha_Z^{-1}(E)$ versus
$t=\ln (E/\Lambda_Z)$ for $m_{\tilde{\bm{\varphi}}_{1}^{(Z)}}
= 300\,GeV$, $m_{\psi^{(Z)}_{2}} = 100\; GeV$ and
$m_{\psi^{(Z)}_{1}}= 1\;MeV$.}
\end{figure}
\begin{figure}
\includegraphics[angle=-90,width=8cm]{alvst_1EV.epsi}
\includegraphics[angle=-90,width=8cm]{alinvst_1EV.epsi} 
\caption{\label{fig5}$\alpha_Z(E)$ and $\alpha_Z^{-1}(E)$ versus
$t=\ln (E/\Lambda_Z)$ for $m_{\tilde{\bm{\varphi}}_{1}^{(Z)}}
= 300\,GeV$, $m_{\psi^{(Z)}_{2}} = 100\; GeV$ and
$m_{\psi^{(Z)}_{1}}= 1\;eV$.}
\end{figure}
\begin{figure}
\includegraphics[angle=-90,width=8cm]{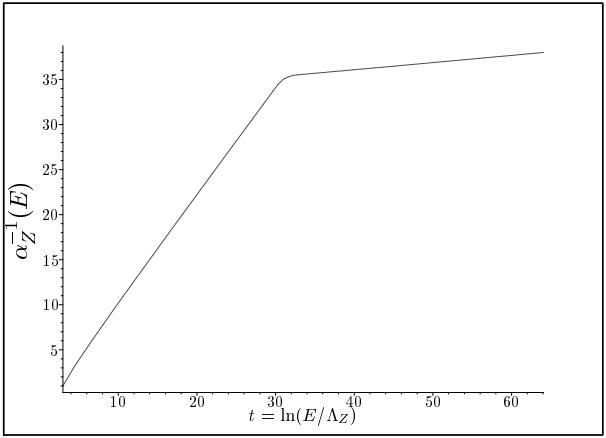} 
\caption{\label{fig6}$\alpha_Z^{-1}(E)$ versus
$t=\ln (E/\Lambda_Z)$ for $m_{\tilde{\bm{\varphi}}_{1}^{(Z)}}
= 300\,GeV$, $m_{\psi^{(Z)}_{2}} = 100\; GeV$ and
$m_{\psi^{(Z)}_{1}}= 50\;GeV$ starting with $\alpha_Z^{-1}(M)=
38$. Here $\alpha_Z^{-1}(E=5.7 \times 10^{-2}\,eV)=1$.}
\end{figure}
\begin{figure}
\includegraphics[angle=-90,width=8cm]{alvst_100.epsi}
\includegraphics[angle=-90,width=8cm]{alinvst_100.epsi} 
\caption{\label{fig7}$\alpha_Z(E)$ and $\alpha_Z^{-1}(E)$ versus
$t=\ln (E/\Lambda_Z)$ for $m_{\tilde{\bm{\varphi}}_{1}^{(Z)}}
= 300\,GeV$, $m_{\psi^{(Z)}_{2}} = 200\; GeV$ and
$m_{\psi^{(Z)}_{1}}= 100\;GeV$.}
\end{figure}
\begin{figure}
\includegraphics[angle=-90,width=8cm]{al200vst_50.epsi}
\includegraphics[angle=-90,width=8cm]{al200invst_50.epsi} 
\caption{\label{fig8}$\alpha_Z(E)$ and $\alpha_Z^{-1}(E)$ versus
$t=\ln (E/\Lambda_Z)$ for $m_{\tilde{\bm{\varphi}}_{1}^{(Z)}}
= 300\,GeV$, $m_{\psi^{(Z)}_{2}} = 200\; GeV$ and
$m_{\psi^{(Z)}_{1}}= 50\;GeV$.}
\end{figure}
\begin{figure}
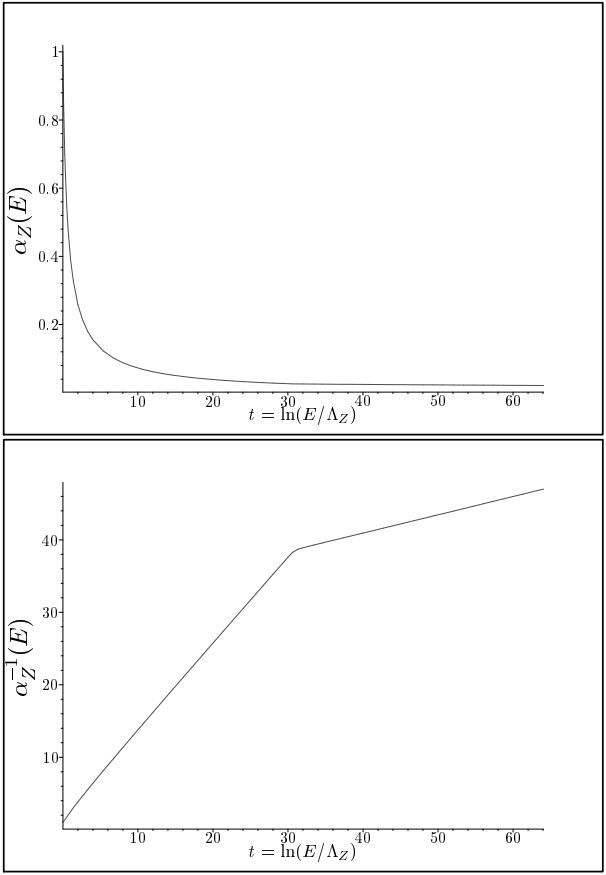

\includegraphics[angle=-90,width=8cm]{doublet_50.epsi}
\includegraphics[angle=-90,width=8cm]{doubletin_50.epsi} 
\caption{\label{fig9}$\alpha_Z(E)$ and $\alpha_Z^{-1}(E)$ versus
$t=\ln (E/\Lambda_Z)$ for the $SU(2)_Z$ doublet
case $m_{\bm{\varphi}_{1}^{(Z)}}
= 300\,GeV$, $m_{\psi^{(Z)}_{2}} = 100\; GeV$ and
$m_{\psi^{(Z)}_{1}}= 50\;GeV$.}
\end{figure}
\begin{figure}
\includegraphics[angle=-90,width=8cm]{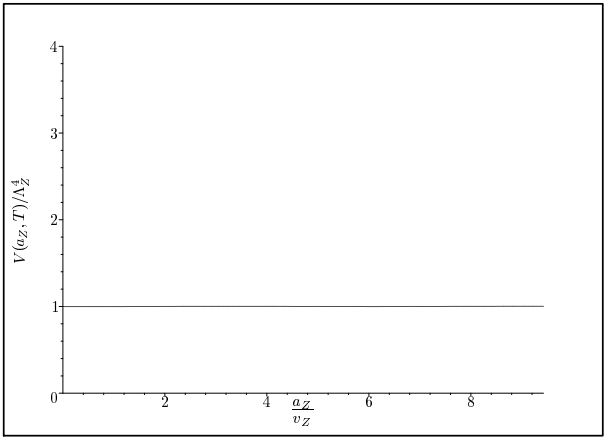}
\caption{\label{fig10}$V(a_Z,T)/\Lambda_Z^4$ as a function
of $a_Z/v_Z$ for $\kappa(T) = 10^{-3}$ with no soft breaking.}
\end{figure}
\begin{figure}
\includegraphics[angle=-90,width=8cm]{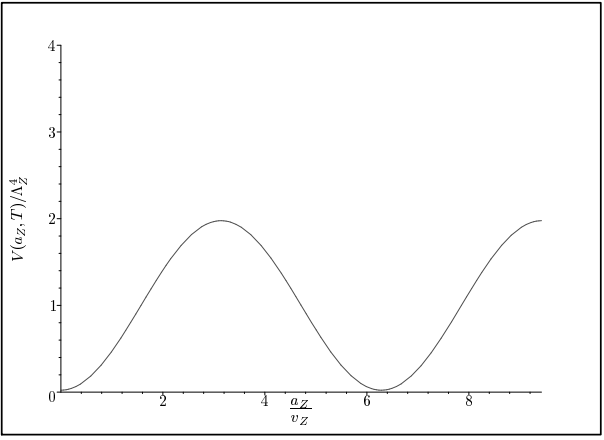}
\caption{\label{fig11}$V(a_Z,T)/\Lambda_Z^4$ as a function
of $a_Z/v_Z$ for $\kappa(T) = 1$ with no soft breaking.}
\end{figure}
\begin{figure}
\includegraphics[angle=-90,width=8cm]{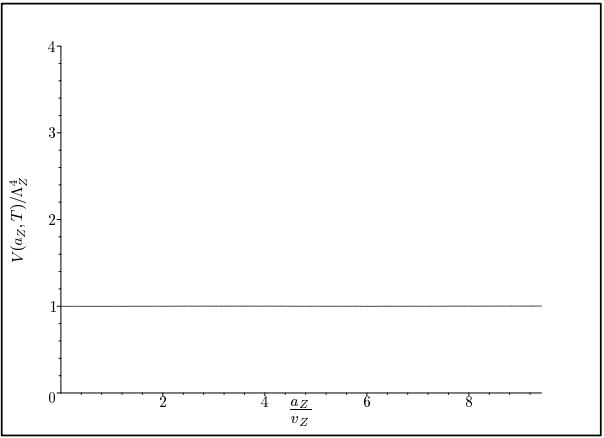}
\caption{\label{fig12}$V(a_Z,T)/\Lambda_Z^4$ as a function
of $a_Z/v_Z$ for $\kappa(T) = 10^{-3}$ with soft breaking.}
\end{figure}
\begin{figure}
\includegraphics[angle=-90,width=8cm]{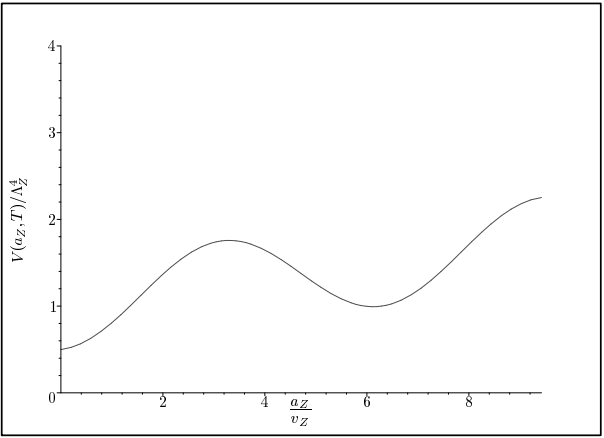}
\caption{\label{fig13}$V(a_Z,T)/\Lambda_Z^4$ as a function
of $a_Z/v_Z$ for $\kappa(T) = 10^{-0.3}$ with soft breaking.}
\end{figure}
\begin{figure}
\includegraphics[angle=-90,width=8cm]{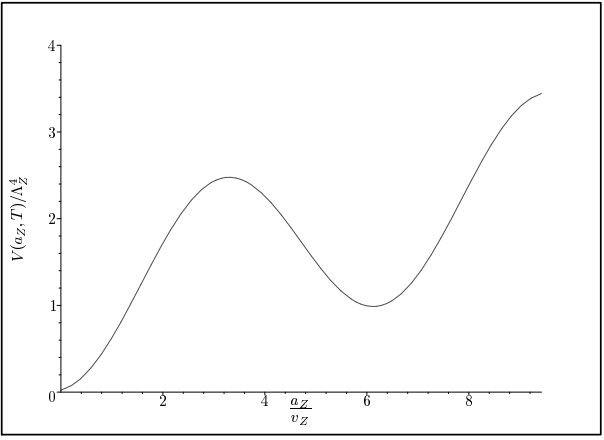}
\caption{\label{fig14}$V(a_Z,T)/\Lambda_Z^4$ as a function
of $a_Z/v_Z$ for $\kappa(T) = 1$ with soft breaking.}
\end{figure}

\begin{thebibliography}{99}
\bibitem{acceleration} S. Permutter {\em et al.}, Astrophys. J. {\bf 517},
565 (1999); A. Riess {\em et al.}, Astron. J. {\bf 116}, 1009 (1998).
\bibitem{riess} A. Riess {\em et al.}, Astrophys. J. {\bf 607}, 665 (2004).
\bibitem{snls} P. Astier {\em et al.}, astro-ph/0510447.
\bibitem{sahni} C. Wetterich, Nucl. Phys. B {\bf 302}, 668 (1988);
B. Ratra and P. J. E. Peebles,  Phys. Rev. D {\bf 37}, 3406
(1988); P. J. E. Peebles and B. Ratra, Astrophys. J. {\bf 325}, L17 (1988);
I. Zlatev, L. Wang, and P. Steinhardt, Phys. Rev. Lett. {\bf 82}, 896 (1999);
 P. Steinhardt, L. Wang, and I. Zlatev, Phys. Rev. D {\bf 59}, 123504 (1999).
See also Varun Sahni, astro-phys/0403324,
for a review and an extensive list of references.
\bibitem{doran} Notice however that there are proposals to detect the influence
of ``Early Dark Energy'' on the Cosmic Microwave Backgound (CMB) as well
as structure formation. See e.g. the following references:
M. Doran, J. Schwindt, and C. Wetterich, Phys. Rev. D {\bf 64}, 123520 (2001);
M. Doran, M. Lilley, J. Schwindt, and C. Wetterich,  Astrophys. J. {\bf 559},
501 (2001); R. Caldwell, M. Doran, C. M\"{u}ller,
G. Sch\"{a}ffer, and C. Wetterich, Astrophys. J. {\bf 591}, L75 (2003);
M. Bartelmann, M. Doran, and C. Wetterich, astro-ph/0507257.
\bibitem{sahlen} M Sahlen, A. Liddle, and D. Parkinson, astro-ph/0507075.
\bibitem{kolb} For a good pedagogical discussion of various aspects
of the false or metastable vacuum and its implications, see
E. W. Kolb and M. S. Turner, {\em The Early Universe},
Addison-Wesley Publishing Company (1990).
\bibitem{su2} P. Q. Hung, hep-ph/0504060.
\bibitem{zophos} Here the subscript $Z$ refers to an
ancient greek word {\em zophos} which means {\em darkness}.
\bibitem{bj} I wish to thank James (bj) Bjorken for asking about
this possibility.
\bibitem{hung2} P. Q. Hung and Paola Mosconi, in preparation.
\bibitem{jain} There exists another proposal to use the 
Peccei-Quinn QCD axion
as an acceleron for the dark energy: Pankaj Jain, Mod. Phys. Lett.
A {\bf 20}, 1763 (2005). (I would like to thank Pankaj Jain for
pointing out this reference.) The axion of our model is however
entirely different from the QCD one used in that paper.
\bibitem{hung3} P. Q. Hung, ``A model of Standard Model leptogenesis'',
in preparation.
\bibitem{PQ} R. Peccei and H. Quinn, Phys. Rev. Lett. {\bf 38}, 1440 (1977).
\bibitem{sikivie} P. Sikivie, Phys. Rev. Lett. {\bf 48}, 1156 (1982).
\bibitem{instanton} E. V. Shuryak, Phys. Lett. B {\bf 79}, 135 (1978);
R. D. Pisarski and L. Yaffe, Phys. Lett. B {\bf 97}, 110 (1980).
\bibitem{sikivie2} See e.g. a nice review by P. Sikivie,
lectures given at 21st Schladming Winter School, Schladming, Austria, 
Feb 26 - Mar 6, 1982.
\bibitem{kamionkowski} For a review, see e.g. K. Kamionkowski, hep-ph/9710467.
\bibitem{kuzmin} V. A. Kuzmin, V. A. Rubakov and M. A. Shaposnikov,
Phys. Lett. B {\bf 155}, 36 (1985).
\bibitem{lepto} M. Fukugita and T. Yanagida, Phys. Lett. B {\bf 174}, 45 (1986);
M. Flanz, E. A. Paschos and U. Sarkar, Phys. Lett. B {\bf 345}, 248 (1995);
L. Covi, E. Roulet and F. Vissani, Phys. Lett. B {\bf 384}, 169 (1996);
W. Buchm\"{u}ller and M. Plumacher, Phys. Lett. B {\bf 431}, 354 (1998).
See also W. Buchm\"{u}ller, R. D. Peccei and T. Yaganida, hep-ph/0502169,
for a review and an extensive list of references.
\bibitem{giudice} G.F. Giudice, A. Notari, M. Raidal, A. Riotto, A. Strumia,
Nucl. Phys. B {\bf 685}, 89 (2004).
\bibitem{abbott} L. F. Abbott, Phys. Lett. B {\bf 150}, 427 (1985).
\end{thebibliography}
\end{document}